\documentclass[aip, jcp, amsmath,amssymb, reprint]{revtex4-1}

\usepackage{dsfont}
\usepackage{lipsum}
\usepackage{xcolor}
\usepackage{enumitem}
\usepackage{graphicx}% Include figure files
\usepackage{dcolumn}% Align table columns on decimal point
\usepackage{bm}% bold math
\usepackage{gensymb}
\usepackage{bbm}

\newcommand{\XP}{\mathbf{x}}       % path vector
\newcommand{\PP}{\mathcal{P}}      % TPS path probability
\newcommand{\DD}{\mathcal{D}}      % path differential
\newcommand{\ZZ}{\mathcal{Z}}

\begin{document}

\title{Combining  multiple interface set path ensembles with MBAR reweighting}

\author{ 
Rik S. Breebaart}
\affiliation{Van 't Hoff Institute for Molecular Sciences, Universiteit van Amsterdam, Science Park 904, 1098 XH Amsterdam, The Netherlands}

\author{Peter G. Bolhuis*}
\affiliation{Van 't Hoff Institute for Molecular Sciences, Universiteit van Amsterdam, Science Park 904, 1098 XH Amsterdam, The Netherlands}
\email{p.gbolhuis@uva.nl}

\begin{abstract}
    We introduce a method to compute the reweighted path ensemble by combining transition interface sampling simulations 
    conditioned on different collective variables.
    The approach is based on the Multistate Bennett Acceptance Ratio (MBAR) methodology applied to entire trajectories. 
     Illustrating the technique with simple 2D potential models and a more complex host-guest system, we show that the statistics can significantly improve compared to a straightforward combination. 
\end{abstract}

\maketitle

\section{Introduction}
 Transition path sampling (TPS) is  a general  simulation technique to study rare complex molecular processes, such as protein conformational changes, nucleation or chemical reactions, introduced originally in  a seminal paper by Dellago et al.~\cite{Dellago1997}. 
In contrast to  enhanced sampling methods that  introduce  a biasing force along a collective variable at the cost of upsetting the natural dynamics\cite{Henin2022},
TPS 
 collects an ensemble of unbiased dynamical pathways that connect initial with final state, by performing Monte Carlo  in trajectory space ~\cite{Dellago1997,Bolhuis2002,Dellago2002,Dellago2009,Bolhuis2009a,BolhuisDellago2015,swensonOpenPathSamplingPythonFramework2019,bolhuis_transition_2021}. 
 Given an initial trajectory the algorithm generates a new trial path by selecting a frame and shooting off a new path by integrating the equations of motions forward and backward in time. The trial path is then accepted with a probability given by the Metropolis acceptance rule, obeying detailed balance.
The resulting transition path ensemble gives  unbiased insight in the mechanism of the rare  transition. 
TPS also gives access to the rate constant  of the rare event via the reversible work between a transition path ensemble and an  ensemble of   trajectories starting in the initial state  but are free to end anywhere.

Notwithstanding the success of TPS,  a major improvement in efficiency was provided by the development of
Transition Interface Sampling (TIS)~\cite{vanErp2003}. This TPS based technique defines a set of hyper-surfaces, or interfaces, in between the stable states, which trajectories need to cross in order to be part of the ensemble~\cite{Bolhuis2009a,vanErp2012,Cabriolu2017}. 
The main objective of TIS is to compute the crossing probability, the probability that a trajectory coming directly from initial state A and crosses an interface, also crosses the next interface. The last interface then yields the crossing probability into the final state B. 
The rate constants follows from
the product over the crossing probabilities for all interfaces,  multiplied by the flux through the first interface. This stratification approach turns an exponential problem into a linear one\cite{vanErp2005,vanErp:2012fe}.
Each interface is parameterised by an order parameter or collective variable (CV) usually denoted $\lambda$. 
TIS yields in principle exact rate constants  independent of the choice of  order parameter, although this choice does influence the sampling statistics. 
Statistics can be further improved by the introduction of replica exchange\cite{vanErp2007,Bolhuis2008,Du2013}, which enhances decorrelation between sampled paths, especially for paths close to the stable states.  
While TIS only needs to compute the crossing probability to the next interface, the crossing probability can be obtained for arbitrary interfaces, and in fact for arbitrary values of $\lambda$. 

 Ref.~\onlinecite{Rogal2010} introduced the reweighted path ensemble (RPE), a conceptual framework arguing that just as configurational samples can  be reweighted to account for the enhanced sampling, using e.g.~umbrella sampling\cite{Torrie_1974} or metadynamics\cite{Laio_2002}, one can also reweight  entire TIS trajectories to account for their being part of a selection-biased ensemble (note that the dynamics is still unbiased). 
The reweighting factors can be obtained applying a  weighted histogram analysis (WHAM)\cite{ferrenbergOptimizedMonteCarlo1989,Kumar1992} to the conditional   crossing probabilities. 
A key insight was that the weight for each trajectory is determined by the maximum $\lambda$ reached by the trajectory\cite{Rogal2010}.
Thus, knowledge of the crossing probability from TIS  leads  to a correct representation of the sampled trajectories in the overall unbiased path ensemble. The advantage is that the RPE gives much more information on paths near the top of the barrier/transition state than an unbiased path ensemble (from straightforward MD) would contain (given the same amount of MD effort). 
In particular, the RPE  gives an accurate projected committor estimate\cite{Lechner2010}.
Reweighted path ensembles  have been computed routinely in several applications, notably  on crystal nucleation \cite{Lechner2011,DazLeines2017,Liang2020,Arjun2020}, membrane permeation\cite{Safaei2025}, protein folding\cite{Du2014,Zhang2024} and association\cite{Newton2015,Newton2017}, as well as in novel algorithms\cite{vanErp2016,breebaart2025understandingreactionmechanismsstart}.
Recently, the interest in the RPE was reinvigorated with the advent of  AI for Molecular Mechanism Discovery (AIMMD) \cite{Jung2023}.
In particular,   Lazzeri  et al extended AIMMD to reconstruct the entire RPE\cite{Lazzeri2023}.

The RPE estimate using WHAM can be  improved by sampling additional interfaces. Also replica exchange TIS helps to improve statistics and convergence\cite{vanErp2007,Bolhuis2008,vanErp:2012fe}. However, one of the limitations of the RPE approach is that the WHAM procedure to obtain the overall total crossing probability  depends on the choice of the CV $\lambda$. If for one reason or the other the interface CV is not optimal, and needs to be changed, e.g. by including additional degrees of freedom,  it is not straightforward to  combine this information with the already existing path ensembles of the previous CV. In fact, one would have to start again and recompute the RPE from scratch.

In this work we show that it is possible to combine multiple path ensembles based on different CVs leading to a general WHAM-like reweighting equation. While it would be possible to develop this WHAM equation by the original approach, we find it clearer to use the  Multistate Bennett Acceptance Ratio  (MBAR) approach\cite{Shirts2008}. 
In the past MBAR and WHAM have been both used for reweighting\cite{Kumar1992,Shirts2008}.  The main difference in perspective is that WHAM is based on the minimization of the statistical error or variance\cite{ferrenbergOptimizedMonteCarlo1989,Kumar1992}, while MBAR can be derived using a likelihood maximization approach\cite{Bartels2000,Shirts2008}. Yet both approaches result in very similar, if not the same, equations. 
We illustrate the novel approach on a simple toy model, and on a more complex host-guest binding transition\cite{breebaart2025understandingreactionmechanismsstart}. 

The remainder of this paper is organized as follows. First, we present the MBAR reweighting procedure for obtaining the RPE from a set of interface ensembles constructed using a single CV interface function, $\lambda$. We then extend this approach to TIS ensembles constructed from two distinct CV functions, $\lambda$ and  $\mu$. Finally, we generalize the method to an arbitrary number $M$ of CV functions. The proposed framework is validated on a simple double-well potential, demonstrating its advantages over applying TIS with a single CV or naively combining the two separately constructed RPEs.

\section{Theory}
We first treat TIS trajectory data in a likelihood framework and show how the RPE arises from a maximum-likelihood  (MBAR) estimator \cite{Bartels2000,Shirts2008,FrenkelSmit} over the biased TIS path ensembles. Concretely, we consider a single interface set $\{\lambda_k\}$ and use TIS to generate trajectory ensembles conditional on crossing each interface $\lambda_k$ in $\lambda$\cite{vanErp2003}. This condition is encoded by the  indicator function $\hat{h}_{A,\lambda}^{k}[\XP]$, which is  unity when the path $\XP$, leaves state A and crosses the interface $\lambda_k$, after which it can return to A or goes on to state B.  Here, a trajectory $\XP=\{x_0,x_1 \dots x_L\}$ consists of $L+1$ frames $x$ separated by time interval $\Delta \tau$, which in turn  contain the position and momenta of all particles in the system. 

Our goal is to estimate path weights $w[\XP]$ and a set of partition sums $\{Z^{k}_{A.\lambda}\}$ that (i) recovers the unbiased ensemble of paths leaving A, $\PP_A[\XP]$, from the union of biased samples and (ii) yields unbiased observables such as the  crossing probabilities $P_{A}(\lambda|\lambda_1)$, with $\lambda_1$ the first interface.
We obtain these factors by maximizing the likelihood of  observing the sampled trajectories under the conditional ensembles, given the unbiased complete path ensemble underlying them, leading to the familiar MBAR fixed-point equation specialized to path space. This one-set formulation establishes notation, clarifies assumptions and exposes the key intuition used later: the weight of a trajectory depends only on the highest interface it crosses. In the next sections, we extend the same likelihood construction to two interface sets made from different collective variables and then to $M$ sets, yielding a unified MultiSet-MBAR procedure for combining trajectories generated with different collective variables as interfaces.

\subsection{RPE based on MBAR for trajectories }
Consider a collection of $K$ trajectory ensembles for the interface set $\{\lambda_{k}\}$, each of $N_k$ samples, with $k=1,\dots,K$.
The unbiased probability $\PP_{A} [\XP]$ to observe a path $\XP$ based on this collection of sampled paths is expressed as 
\begin{align}
    \PP_{A}[\XP] = \frac{1}{\mathcal{Z}} \sum_k^K \sum_n^{N_k}  w_{k,n} \delta[\XP-\XP_{k,n}].
    \label{eq:px}
\end{align}
Here, index $k$ runs over $K$ TIS  simulations, while $n$ runs over the $N_k$ samples in each interface simulation.  
The  factor $\mathcal{Z} = \sum_k \sum_n w_{k,n} $ normalizes the distribution. 
The function $\delta[\XP-\XP_{k,n}]$ returns unity when the trajectory $\XP$ and $\XP_{k,n}$ are identical, and zero otherwise.
The weights of each path $w_{k,n}\equiv w[\XP_{k,n}]$ are unknown, but are also related to the biased path ensembles. 
For instance, in TIS we  bias path ensembles by imposing at each interface  a indicator function $\hat{h}_{A,\lambda}^k[\XP]$, which is only unity if the sampled path from A crosses interface $\lambda_k$.     Combining all $K$ simulations, the distribution   $\PP^{k'}_{A,\lambda}[\XP]$ for such a biased ensemble is 
\begin{align}
    \PP^{k'}_{A,\lambda}[\XP] = \frac{1}{\mathcal{Z}^{k'}_{A,\lambda}}   \sum_{k}^{K} \sum_n^{N_k} w_{k,n} \hat{h}_{A,\lambda}^{k'}[\XP]  \delta[\XP-\XP_{k,n}],
    \label{eq:pk}
\end{align}
with  $\mathcal{Z}^{k'}_{A,\lambda} =  \sum_{k}^{K}\sum_n^{N_k}  w_{k,n} \hat{h}_{A,\lambda}^{k'}[\XP_{k,n}]$.  Note that all paths from all simulations contribute to this distribution.

In MBAR the unknown weights $w_{k,n}$ are determined from the likelihood to observe the sampled paths in its conditional ensemble. The log-likelihood is expressed as
\begin{align}
    \ln L = \ln \prod_{k}^K \prod_{n}^{N_k} \PP^k_{A,\lambda}[\XP_{k,n}] = \sum_{k}^K \sum_{n}^{N_k}  \ln\PP^k_{A,\lambda}[\XP_{k,n}]. 
\end{align}
Substituting $\PP^k_{A,\lambda}[\XP_{k,n}] $ from Eq.\ref{eq:pk} gives
\begin{align}
    \ln L = \sum_{k}^K \sum_{n}^{N_k}  \ln  
    \frac{w_{k,n} \hat{h}_{A,\lambda}^k[\XP_{k,n}] }{\mathcal{Z}^k_{A,\lambda}} .
\end{align}
Optimizing for  $w_{j,n}$ gives 
\begin{align}
    \frac{\partial \ln L}{\partial  w_{j,n} } = 0 = \frac{1}{w_{j,n}} - \sum_{k}^K  \frac{N_k  \hat{h}_{A,\lambda}^k[\XP_{j,n}]} {\mathcal{Z}^k_{A,\lambda}},
    \label{eq:singleSetLoglikelihood}
\end{align}
where the factor $N_k$ derives from the summation over $n$.
Rearranging and dividing by $\mathcal{Z}$  gives 
\begin{align}
    \frac{w_{j,n}}{\mathcal{Z}} = \left[ \sum_{k}^K  \frac{N_k  \hat{h}_{A,\lambda}^k[\XP_{j,n}]} {\mathcal{Z}^k_{A,\lambda}/ \mathcal{Z}}\right]^{-1}.
\end{align}
Inserting this result in the definition of $\mathcal{Z}^{k'}_{A,\lambda}$, Eq.~\ref{eq:pk}, yields 
\begin{align}
    \frac{Z^{k'}_{A,\lambda}}{\mathcal{Z}} = 
    %\sum_{j}^K 
   \sum_j^K \sum_{n}^{N_j}   \frac{\hat{h}_{A,\lambda}^{k'}[\XP_{j,n}]} { \sum_{k}^K  \frac{N_k  \hat{h}_{A,\lambda}^k[\XP_{j,n}]} {\mathcal{Z}^k_{A,\lambda}/ \mathcal{Z}}},
    \label{eq:MBARWHAMeq}
\end{align}
where we for clarity replaced the outer $k$ index by $j$.
This is the implicit function that is central in MBAR\cite{Shirts2008}, and which is identical to the central equation in WHAM\cite{ferrenbergOptimizedMonteCarlo1989} and can be solved iteratively.

With the weights obtained through iteratively solving Eq.~\ref{eq:MBARWHAMeq},
%setting $\mathcal{Z}_1$, 
we can also obtain the unbiased probability from Eq.~\ref{eq:px}
\begin{align}
    \PP_{A}[\XP] = \sum_j^K \sum_n^{N_j} 
   \frac{  \delta[\XP-\XP_{j,n}]}  { \sum_{k}^K  \frac{N_k  \hat{h}_{A,\lambda}^k[\XP_{j,n}]} {\mathcal{Z}^k_{A,\lambda}/ \mathcal{Z}}}.
\end{align}

For TIS we know that the sum in the denominator only has contributions for paths where $\hat{h}_A^k[\XP_{k,n}]$ is nonzero. Denoting the maximum interface that a path crosses as $k_{\rm max}$ (defined by $\lambda^{\rm max}[\XP_{j,n}]$), the denominator reduces to 
\begin{align}
w_{j,n} =    \left( \sum_{k}^{k_{\rm max}[\XP_{j,n}]}  \frac{N_k\mathcal{Z}}  {\mathcal{Z}^k_{A,\lambda}} \right)^{-1},
\label{eq:SingleSetMBARWeights}
\end{align}
which looks very similar to the WHAM expression for the path weights in  Ref.~\cite{Rogal2010}. This means that the weight of each path is determined by the  maximum interface it crosses, as given by $\lambda^{\rm max}[\XP_{j,n}]$. Indeed, this the central result from the RPE treatment\cite{Rogal2010}. The unbiased  path probability is   then given by   Eq.~\ref{eq:px}.

Of particular interest is also the total crossing probability constructed from all the reweighted paths:
\begin{align}    
P_A(\lambda|\lambda_1) 
&= \sum_j^K \sum_{n}^{N_j} w_{j,n} { \theta( \lambda^{\rm max} [\XP_{j,n}]- \lambda) }.
\label{eq:crosproball}
\end{align}
The crossing probability  at the last interface $\lambda_B=\lambda_K$ is 
\begin{align}    
P_A(\lambda_B|\lambda_1) 
= n_{AB} w_{AB},
\end{align}
where 
$w_{AB} =     (\sum_{k}^{K}  \frac{N_k\mathcal{Z}}  {\mathcal{Z}^k_{A,\lambda}})^{-1} $ is the weight for the reactive paths, and  $n_{AB}$ is the cumulative number of reactive paths sampled in the interface ensembles. 

Besides the AB transition we can also sample the reverse or backward BA transition using TIS, resulting in its own RPE, with  weights $w^B[\XP]$. Similarly, we denote the AB transition weights as $w^A[\XP]$. 

From the computed weights $w_{j,n}$ we can construct the  total reweighted path ensemble (RPE). The weight for each trajectory ${\bf x}_{j,n}$ in the sampled  path ensemble starting from A is
\begin{equation}
    w^A[{\bf x}_{j,n}] = c_A  {w}^A_{j,n}. 
    \label{eq:weights_RPE_A}
\end{equation}

Likewise, the weight for each trajectory ${\bf x}_{j,n}$ in the sampled path ensemble starting from B is
\begin{equation}
    w^B[{\bf x}_{j,n}] = c_B  {w}^B_{j,n},
    \label{eq:weights_RPE_B}
\end{equation}
where the labels A and B,  distinguish the forward and backward transitions, respectively.
The unknown constants $c_A$ and $c_B$ are determined by matching the forward and reverse crossing probability for the  reactive paths (e.g. by having the (flux) contribution of the reactive path weights $n_{AB} w_{AB}=n_{BA} w_{BA}$ be equal for forward and reverse trajectories.)

The final total RPE weight for each sampled path is then simply the sum $\mathbbm{1}_A(x_0 ) w^A[{\bf x}] + \mathbbm{1}_B(x_0)  w^B[{\bf x}]$, where the stable state characteristic functions $\mathbbm{1}_A(x)$ return unity when the configuration is in the state $A$ definition, and zero otherwise.  
This sum, however, only takes into account paths that leave A (or B) and cross any of the interfaces, not those paths that remain in the stable state A (or B). However, these paths can easily be added by simulations in the stable states, which are then weighted by  matching  their flux through the first interface or by employing the minus or plus ensembles (in RETIS\cite{vanerpReactionRateCalculation2007,bolhuisRareEventsMultiple2008}). 

\subsection{RPE based on MBAR for two TIS interface sets}
\label{sec:twosets}

Imagine now that we have two sets of TIS simulations, respectively with $K_1$ and $K_2$ interfaces. The first interface set $\{\lambda_k\}$ is based on a function $\lambda(x)$, the second interface set $\{\mu\}$ is based  on  $\mu(x)$, which might be  a different function. We assume that the stable state definitions are identical, i.e. $\lambda_0\equiv\mu_0$, but that  all other interfaces $i>0$ are  described by different functions $\lambda(x) \neq \mu(x)$.
From these TIS simulations we obtain two sets of paths $\XP_{k,n}$ and $\tilde{\XP}_{k,n}$, where $\XP_{k,n}$ is the $n$th path sampled on the $k$th interface of the first set, conditioned according to the $\lambda$ interfaces. Similarly,  $\tilde{\XP}_{k,n}$ denotes the $n$th path sampled on the $k$th interface of the second set,  which biases  according to the $\mu$ interfaces.
Here, as before, indexes $k$ run over $K_{1,2}$ interface simulations, while $n$ runs over the $N_k$ samples in each simulation.  
The idea is that we can join the two sets with the probability that we observe a path  $\XP$ in an unbiased ensemble as 
\begin{align}
    \PP_{A}[\XP] &= \frac{1}{\mathcal{Z}} \left( \sum_k^{K_1} \sum_n^{N_k}  w_{k,n} \delta[\XP-\XP_{k,n}]  \right. \notag \\ & +\left.  \sum_k^{K_2} \sum_n^{\tilde{N}_k}  \tilde w_{k,n} \delta[\XP-\tilde \XP_{k,n}] \right),
    \label{eq:twosetUnbiased}
\end{align}
where the factor  
\begin{align}
\mathcal{Z} = \sum_k^{K_1} \sum_n^{N_k} w_{k,n} +  \sum_k^{K_2} \sum_n^{{N}_k} \tilde w_{k,n},
\end{align}
normalizes the distribution.

The weights of each path $w_{k,n}$ and $\tilde w_{k,n}$ are unknown, but are related to the biased TIS path ensembles they are sampled from. Just as in the previous section the distribution of paths which satisfy the condition of crossing the $k$th interface function based on $\lambda$, given by indicator function $\hat{h}_{A\lambda}^k[\XP]$, is:
\begin{align}
    \PP^{k'}_{A,\lambda}[\XP] 
 = \frac{1}{\mathcal{Z}^{k'}_{A,\lambda}}  \left(  \sum_{k}^{K_1} \sum_n^{N_k} w_{k,n} \hat{h}_{A,\lambda}^{k'}[\XP]  \delta[\XP-\XP_{k,n}]
    \right. \notag \\  + \left.\sum_{k}^{K_2} \sum_n^{\tilde{N}_k} \tilde w_{k,n} \hat{h}_{A,\lambda}^{k'}[ \XP]  \delta[\tilde \XP-\tilde \XP_{k,n}]
    \right).
    \label{eq:PAlambda}
\end{align}
Note that  sampled paths from all $K_1 + K_2$  TIS simulations can contribute to this  distribution. 
The conditional partition function is given by
\begin{align}
\mathcal{Z}^{k'}_{A,\lambda} =  \sum_{k}^{K_1}\sum_n^{N_k}  w_{k,n} \hat{h}_{A,\lambda}^{k'}[\XP_{k,n}]  + 
  \sum_{k}^{K_2}\sum_n^{\tilde{N}_k}  \tilde w_{k,n} \hat{h}_{A,\lambda}^{k'}[\tilde \XP_{k,n}].
\end{align}

We can look at a similar distribution of paths for the interfaces based on the $\mu$ function, using $\hat{h}_{A,\mu}^{k}[\XP]$ as an indicator   function:
\begin{align}
    \PP^{k'}_{A,\mu}[\XP] 
 &= \frac{1}{\mathcal{Z}^{k'}_{A,\mu}}  \left(  \sum_{k}^{K} \sum_n^{N_k} w_{k,n} \hat{h}_{A,\mu}^{k'}[\XP]  \delta[\XP-\XP_{k,n}]
    \right. \notag \\  &+ \left.\sum_{k}^{K} \sum_n^{\tilde{N}_k} \tilde w_{k,n} \hat{h}_{A,\mu}^{k'}[ \XP]  \delta[\tilde \XP-\tilde \XP_{k,n}]
    \right),
      \label{eq:PAmu}
\end{align}
where 
\begin{align}
\mathcal{Z}^{k'}_{A,\mu} =  \sum_{k}^{K_1}\sum_n^{N_k}  w_{k,n} \hat{h}_{A,\mu}^{k'}[\XP_{k,n}]  + 
  \sum_{k}^{K_2}\sum_n^{\tilde{N}_k}  \tilde w_{k,n} \hat{h}_{A,\mu}^{k'}[\tilde \XP_{k,n}].
\end{align}
Since the paths $\XP_{k,n}$ for $n\in 1,\ldots N_k$ are sampled using the  function $\lambda(x)$ at $\lambda_k$, the corresponding weights $w_{k,n}$ must be determined based on the likelihood of these paths under the conditional distribution $\PP^{k}_{A,\lambda}[\XP]$. Similarly, the paths $\tilde{\XP}_{k,n}$, which are sampled according to $\mu$ at $\mu_k$, are drawn from the conditional distribution $\tilde{\PP}^{k}_{A,\mu}[\XP]$. 
However, the partition sums $Z^{k}_{A,\lambda}$ and $\tilde{Z}^{k}_{A,\mu}$ are computed using both sets of paths. This joint construction allows for additional information about the unbiased distribution to be extracted from the combined datasets, thereby improving the estimation of the partition sums.

The likelihood of the paths $\XP$ drawn from the $k$th $\lambda$-biased ensemble is given by 
\begin{equation}
    L_{k,\lambda}(\XP|\{w_{k,n}\},\{\tilde{w}_{k,n}\}) = \PP^{k}_{A,\lambda}[\XP],
\end{equation}
and for the $\tilde{\XP}$ paths it is 
\begin{equation}
    L_{k,\mu}(\tilde\XP|\{w_{k,n}\},\{\tilde{w}_{k,n}\}) = \PP^{k}_{A,\mu}[\tilde\XP],
\end{equation}
based on $\mu$-biased data.
The log-likelihood of the observed data of  both sets is thus
\begin{align}
    \ln L =& \ln \prod_{k}^{K_1} \prod_{n}^{N_k} \PP_{A,\lambda}^k[\XP_{k,n}]  \prod_{k}^{K_2} \prod_{n}^{\tilde{N}_k} \PP_{A,\mu}^k[\tilde \XP_{k,n}] \notag \\
    =& \sum_{k}^{K_1} \sum_{n}^{N_k}  \ln\PP_{A,\lambda}^k[\XP_{k,n}]  +  \sum_{k}^{K_2} \sum_{n}^{\tilde{N}_k}  \ln\PP_{A,\mu}^k[\tilde \XP_{k,n}].
\end{align}
Substitution of the distributions \ref{eq:PAlambda} and \ref{eq:PAmu} gives 
\begin{align}
    \ln L = \sum_{k}^{K_1} \sum_{n}^{N_k} ( \ln  w_{k,n} \hat{h}_{A,\lambda}^{k}[\XP_{k,n}]  - \ln  {Z}^{k}_{A,\lambda}) \notag \\+
    \sum_{k}^{K_2} \sum_{n}^{\tilde{N}_k} ( \ln \tilde w_{k,n} \hat{h}_{A,\mu}^{k}[\tilde \XP_{k,n}]  - \ln  {Z}^{k}_{A,\mu}).
\end{align}
Maximizing this w.r.t. $w_{j,n}$ and $\tilde{w}_{j,n}$ gives 
\begin{align}
    \frac{\partial \ln L}{\partial  w_{j,n} } = 0 = \frac{1}{w_{j,n}} - \sum_{k}^{K_1}  \frac{N_k  \hat{h}_{A,\lambda}^k[\XP_{j,n}]} {\mathcal{Z}_{A,\lambda}^{k}} - \sum_{k}^{K_2}  \frac{\tilde{N}_k  \hat{h}_{A,\mu}^k[\XP_{j,n}]} {\mathcal{Z}_{A,\mu}^{k}}
    \label{eq:log_max_lambda_weight}
\end{align}
and 
\begin{align}
    \frac{\partial \ln L}{\partial  \tilde{w}_{j,n} } = 0 = \frac{1}{\tilde{w}_{j,n}} -  \sum_{k}^{K_1}  \frac{N_k  \hat{h}_{A,\lambda}^k[\tilde{\XP}_{j,n}]} {\mathcal{Z}_{A,\lambda}^{k}} - \sum_{k}^{K_2}  \frac{\tilde{N}_k  \hat{h}_{A,\mu}^k[\tilde{\XP}_{j,n}]} {\mathcal{Z}_{A,\mu}^{k}}
    \label{eq:log_max_mu_weight}
\end{align}

Rearranging expression eq.(\ref{eq:log_max_lambda_weight}) and eq.(\ref{eq:log_max_mu_weight}) gives:
\begin{align}
    \frac{w_{j,n} }{\ZZ}= \left[\sum_{k}^{K_1}  \frac{N_k  \hat{h}_{A,\lambda}^k[\XP_{j,n}]} {\mathcal{Z}_{A,\lambda}^{k}/\ZZ} +\sum_{k}^{K_2}  \frac{\tilde{N}_k  \hat{h}_{A,\mu}^k[\XP_{j,n}]} {\mathcal{Z}_{A,\mu}^{k}/\ZZ}\right]^{-1}
    \label{eq:wjn_twoset}
\end{align}
and 
\begin{align}
    \frac{\tilde{w}_{j,n} }{\ZZ} = \left[\sum_{k}^{K_1}  \frac{N_k  \hat{h}_{A,\lambda}^k[\tilde{\XP}_{j,n}]} {\mathcal{Z}_{A,\lambda}^{k}/\ZZ} + \sum_{k}^{K_2}  \frac{\tilde{N}_k  \hat{h}_{A,\mu}^k[\tilde{\XP}_{j,n}]} {\mathcal{Z}_{A,\mu}^{k}/\ZZ}\right]^{-1},
\end{align}
which are actually identical. 
Using these weights in the expression for the conditional partition functions we get 
\begin{align}
    \frac{\ZZ^{k'}_{A,\lambda}}{\ZZ} &=  \sum_{j}^{K_1}\sum_n^{N_j}  \frac{{h}_{A,\lambda}^{k'}[\XP_{j,n}]}{\sum_{k}^{K_1}  \frac{N_k  \hat{h}_{A,\lambda}^k[\XP_{j,n}]} {\ZZ_{A,\lambda}^{k}/\ZZ} +\sum_{k}^{K_2}  \frac{\tilde{N}_k  \hat{h}_{A,\mu}^k[\XP_{j,n}]} {\ZZ_{A,\mu}^{k}/\ZZ} } \notag \\ &+
    \sum_{j}^{K_2}\sum_n^{\tilde N_j} \frac{{h}_{A,\lambda}^{k'}[\tilde \XP_{j,n}] }{\sum_{k}^{K_1}  \frac{N_k  \hat{h}_{A,\lambda}^k[\tilde{\XP}_{j,n}]} {\mathcal{Z}_{A,\lambda}^{k}/\ZZ} + \sum_{k}^{K_2}  \frac{\tilde{N}_k  \hat{h}_{A,\mu}^k[\tilde{\XP}_{j,n}]} {\mathcal{Z}_{A,\mu}^{k}/\ZZ} },
\end{align}
and similarly 
\begin{align}
    \frac{\ZZ^{k'}_{A,\mu}}{\ZZ} =&  \sum_{j}^{K_1}\sum_n^{N_j}  \frac{\hat{h}_{A,\mu}^{k'}[\XP_{j,n}]}{\sum_{k}^{K_1}  \frac{N_k  \hat{h}_{A,\lambda}^k[\XP_{j,n}]} {\ZZ_{A,\lambda}^{k}/\ZZ} +\sum_{k}^{K_2}  \frac{\tilde{N}_k  \hat{h}_{A,\mu}^k[\XP_{j,n}]} {\ZZ_{A,\mu}^{k}/\ZZ} } \notag\\ &+
    \sum_{j}^{K_2}\sum_n^{\tilde N_j} \frac{ \hat{h}_{A,\mu}^{k'}[\tilde \XP_{j,n}] }{\sum_{k}^{K_1}  \frac{N_k  \hat{h}_{A,\lambda}^k[\tilde{\XP}_{j,n}]} {\mathcal{Z}_{A,\lambda}^{k}/\ZZ} + \sum_{k}^{K_2}  \frac{\tilde{N}_k  \hat{h}_{A,\mu}^k[\tilde{\XP}_{j,n}]} {\mathcal{Z}_{A,\mu}^{k}/\ZZ}},
\end{align}
which can be solved iteratively, leading to a MBAR procedure for two TIS sets obtained through different interface functions. The method gives the same solution as original MBAR for a single set when the interface function of the two sets is identical (see Appendix) .

\subsubsection*{Iteratively solving the two set MBAR equation}
With initial values  $\ZZ^{k'}_{A,(\lambda/\mu)}=1$, the above set of coupled equations can be solved iteratively for ${\ZZ^{k'}_{A,(\lambda/\mu)}}/{\ZZ}$  
in order to obtain the weights for each path. Similar to standard MBAR (or WHAM), the denominator in the weight expressions only receives contribution from paths for which $\hat{h}_{A(\lambda/\mu)}^{k}[\XP]\neq0$. Denoting the maximum interface crossed in the $\lambda$ and $\mu$ CVs as $k_{\rm max}^{\lambda}$ and $k_{\rm max}^{\mu}$, respectively, the path weights in Eq.~\ref{eq:wjn_twoset} become:
\begin{equation}
    \begin{split}
       w[\XP_{j,n}] 
        =& \left[\sum_{k}^{k^{\lambda}_{\rm max}[\XP_{j,n}]}  \frac{N_k } {\ZZ_{A,\lambda}^{k}/\ZZ} +\sum_{k}^{k^{\mu}_{\rm max}[\XP_{j,n}]}  \frac{\tilde{N}_k  } {\ZZ_{A,\mu}^{k}/\ZZ}\right]^{-1}.
    \end{split}
\label{eq:MBARtwoSetWeight}
\end{equation}

This means that the weight of each path is determined by the maximum it reaches in both $\lambda$ and $\mu$ interfaces. Thus, keeping track of the maximum of each path allows us to compute the weights of each trajectory both for the $\lambda$ and $\mu$ sampled trajectories.
Using the weights of Eq.~\ref{eq:MBARtwoSetWeight} for a given ${\ZZ^{k'}_{A,(\lambda/\mu)}}/{\ZZ}$ we can recompute the partition sums:
\begin{eqnarray}
        \frac{\mathcal{Z}^{k'}_{A,\lambda}}{\mathcal{Z}} =  \sum_{j}^{K_1}\sum_n^{N_j}  \hat{h}_{A,\lambda}^{k'}[\mathbf{x}_{j,n}] w[\mathbf{x}_{j,n}]  +
    \sum_{j}^{K_2}\sum_n^{\tilde N_j} \hat{h}_{A,\lambda}^{k'}[\tilde{\mathbf{x}}_{j,n}] w[\tilde{\mathbf{x}}_{j,n}] \notag \\
    \frac{\mathcal{Z}^{k'}_{A,\mu}}{\mathcal{Z}} =  \sum_{j}^{K_1}\sum_n^{N_j}  \hat{h}_{A,\mu}^{k'}[\mathbf{x}_{j,n}] w[\mathbf{x}_{j,n}]  +
    \sum_{j}^{K_2}\sum_n^{\tilde N_j} \hat{h}_{A,\mu}^{k'}[\tilde{\mathbf{x}}_{j,n}] w[\tilde{\mathbf{x}}_{j,n}] , \notag   
    \label{eq:paritionterms_two_set}
\end{eqnarray}
and repeat  this for  both sets and interfaces until convergence.

\subsubsection*{Crossing probabilities}
The resulting  crossing probabilities \(P_A(\lambda|\lambda_1)\) can now be constructed from the full, reweighted ensemble of trajectories across both \(\lambda\) and \(\mu\) sets:

\begin{equation} 
    \begin{split}
        P_A(\lambda|\lambda_1) &=\sum_k^{K_{\lambda}} \sum_n^{N_k}   \theta( \lambda^{\text{\rm max}} [\XP_{k,n}]- \lambda)  w^A[\XP_{k,n}]  \hat{h}_{A,\lambda_1}[\XP_{k,n}] \\
        & + \sum_k^{K_{\mu}} \sum_n^{\tilde{N}_k} \theta( \lambda^{\text{\rm max}} [\tilde \XP_{k,n}]- \lambda)  w^A[\tilde \XP_{k,n}]  \hat{h}_{A,\lambda_1}[\tilde\XP_{k,n}], 
        \label{eq:crosprob_twoSet}
    \end{split}
\end{equation}

Here, the indicator function \(h_{A,\lambda_1}[\XP]\) explicitly enforces that the path has crossed the reference interface \(\lambda_1\), ensuring correct conditioning of the crossing probability. While this condition is always satisfied implicitly in standard TIS based on a single interface set $\{\lambda_k\}$  as all paths go beyond \(\lambda_1\), in the two-set MBAR formulation it must be included explicitly, especially when one of the ensembles originates from a different CV or contains interfaces preceding \(\lambda_1\). A similar expression holds for \(P_A(\mu|\mu_1)\).

\subsubsection*{Construct RPE from Two sets}
\label{sec:RPEconstruction}

From the computed two-set weights using Eq.\ref{eq:MBARtwoSetWeight} for both the forward ($A\rightarrow B$) and backward ($B\rightarrow A$) transitions, we can construct the total RPE in a manner analogous to the single-set scheme. Each trajectory starting in A (from both the $\lambda$ and $\mu$ set) is assigned a weight and is scaled with a constant factor $c_A$ for the forward ensembles as in Eq.~\ref{eq:weights_RPE_A}. Similarly, trajectories starting in B are weighted according to the backward ensembles and scaled with a factor $c_B$ as in Eq~\ref{eq:weights_RPE_B}. We then tune the unknown constants $c_A$ and $c_B$ to ensure that the weighted flux contribution of the reactive paths is equal for the forward and reverse process.

\begin{figure}
    \centering
    \includegraphics[width=\linewidth]{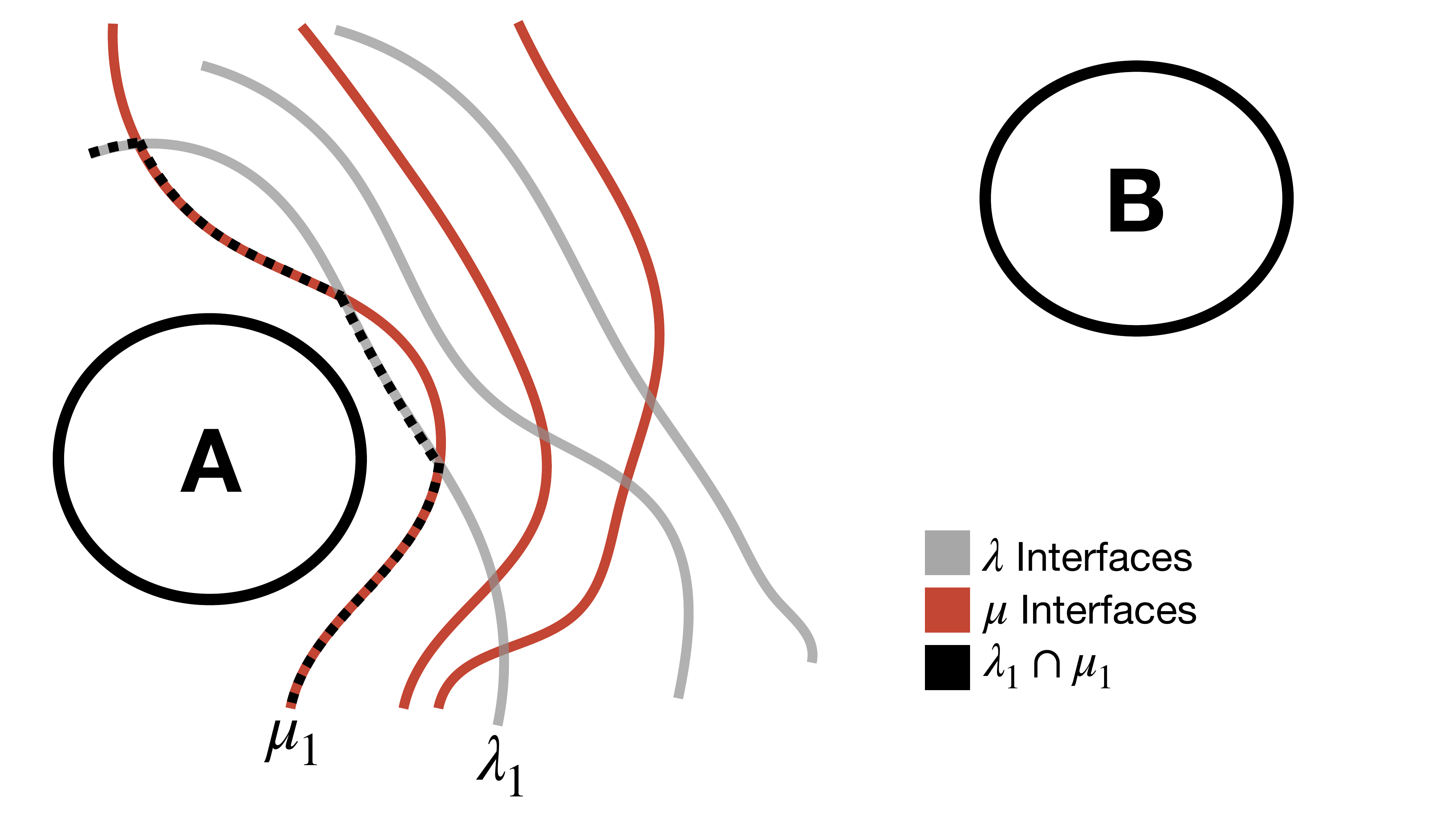}
    \caption{Illustration of the intersection of $\lambda$ and $\mu$ interfaces and the joint initial interface defined as the first interface that trajectories cross. }
    \label{fig:Interface_schematic}
\end{figure}
The RPE constructed via TIS does not intrinsically contain information about the stable state populations. As in the single-set situation, stable state information can be introduced post hoc by simulating the stable states and matching the flux through the first interface on which the TIS ensembles have been conditionally sampled\cite{Rogal2010,vanErp2016}.
In the case of two-sets, the flux can be matched through the intersection of the first interfaces of the $\lambda$ and $\mu$ sets, specifically the region where both interface definitions first overlap. This matching interface represents the entry point to the conditional ensembles and serves as the region beyond which included trajectories in the TIS ensembles have been conditionally sampled.

The matching is performed at the intersection of the first interfaces ($\lambda_1\cap \mu_1$), illustrated by the black dotted line in Fig.~\ref{fig:Interface_schematic}. This intersection defines the part of phase space located outside the region in which  any conditional sampling is applied. All trajectories included in the RPE cross this interface, and as such, it provides natural basis for aligning the stable state flux.

The partition function $\ZZ_A$ of all paths from A that belong to the two-set RPE ensembles and thus those proceeding past the intersecting interface represents the total weight of all paths that enter the conditionally sampled regions (so all paths coming from the TIS simulations).
In matching the flux of the stable state simulations to those from the TIS RPE we ensure that $n_{A,\lambda_1\cap\mu_1}w_{\rm ss}^A=\ZZ_A$,where $w_{\rm ss}^A$ is a weight assigned to the stable state trajectories to ensure the flux is matched, and $n_{A,\lambda_1\cap\mu_1}$ is the amount of effective first interface crossings through the joint interfaces. For simplicity the parts of the stable state trajectories beyond the $\lambda_1\cap\mu_1$ interface are removed and the part of the trajectories from the RPE before crossing $\lambda_1\cap\mu_1$ are removed. 

\subsection{Extension to M interface sets}

The derivation presented for the two-set case can be directly generalized to an arbitrary number of interface ensemble sets $M$. Each set $i$ is defined by its own interface function $\lambda^{(i)}$ with $K^{(i)}$ interfaces and corresponding conditional partition functions $\mathcal{Z}_{A,\lambda^{(i)}}^{k}$. Assuming all sets share the same stable states ($\lambda_A \equiv \lambda_0^{(i)}$), the joint likelihood over all sampled paths can be maximized to obtain a self-consistent expression for the unbiased path weights.  
This yields for each path $\mathbf{x}$ the weight  
\begin{align}
w^{A}[\mathbf{x}] = 
\left[
\sum_{i=1}^{M}
\sum_{k}^{K^{(i)}}
\frac{N^{(i)}_k\,\hat{h}_{A,\lambda^{(i)}}^{k}[\mathbf{x}]}
{\mathcal{Z}_{A,\lambda^{(i)}}^{k}/\mathcal{Z}}
\right]^{-1},
\end{align}
where $N^{(i)}_k$ denotes the number of sampled paths in interface ensemble $k$ of set $i$, and $\hat{h}_{A,\lambda^{(i)}}^{k}[\mathbf{x}]$ is the indicator function that evaluates to unity if the path satisfies the corresponding interface condition.  

Because $\hat{h}_{A,\lambda^{(i)}}^{k}[\mathbf{x}] = 0$ for interfaces beyond the maximal crossing of $\mathbf{x}$ along $\lambda^{(i)}$, the weight depends only on $k^{(i)}_{\rm max}$,   the highest interface reached in each set:
\begin{align}
w^{A}[\mathbf{x}] =
\left[
\sum_{i=1}^{M}
\sum_{k}^{k_{\text{\rm max}}^{(i)}[\mathbf{x}]}
\frac{N^{(i)}_k}
{\mathcal{Z}_{A,\lambda^{(i)}}^{k}/\mathcal{Z}}
\right]^{-1}.
\label{eq:multiset_MBAR_weight}
\end{align}
This expression generalizes the two-set result and preserves the desirable normalization and reweighting properties of MBAR across all $M$ interface ensemble sets.  The MultiSet-MBAR procedure recovers Eq.~\ref{eq:SingleSetMBARWeights} for $M=1$ and Eq.~\ref{eq:MBARtwoSetWeight} if $M=2$. The weights to reconstruct the equilibrium distribution of the paths only depend on the maximum interface crossed in each set.
The iterative solution for the partition functions $\mathcal{Z}_{A,\lambda^{(i)}}^{k}$ are then
\begin{equation}
    \frac{\mathcal{Z}_{A,\lambda^{(i)}}^{k'}}{\ZZ} = \sum^{M}_{i=1}\sum_{j}^{K^{(i)}}\sum_n^{N^{(i)}_j}  \hat{h}_{A,\lambda^{(i)}}^{k'}[\XP^{(i)}_{j,n}]
    w^A[\XP^{(i)}_{j,n}]   
\end{equation}
A detailed derivation of these expressions is provided in the Appendix section~\ref{A:MultiSetMBAR_Derivation}.

\section{Results}

\subsection{The 2D double well}
We tested the MBAR reweighting procedure on a 2D double well potential, given by a one dimensional double well potential in $x$ with an additional harmonic potential in $y$:
\begin{equation}
U_{\rm dw}(x,y) = A(x^2-x_0^2)^2 + \omega y^2,
\end{equation} with $A=1$ and $x_0=1$ and $\omega=1$, resulting in a saddle point at $x=0$ and two minima at $x=\pm1$. For the TIS simulations the stable states of the system are defined as $x<-0.9$ for the A state and $x>0.9$ for the B state.
For the dynamics we employed a Langevin BOAOB integrator \cite{leimkuhlerRationalConstructionStochastic2012} in the   OpenPathSampling (OPS) package\cite{swensonOpenPathSamplingPythonFramework2019} with the temperature set to $T_0=0.1$, the friction $\gamma = 10$ and a time step $dt=0.05$, yielding a barrier height of $1/T_0$ in units of the thermal energy $k_B T$, with $k_B$ the Boltzmann constant.

Straightforward Langevin simulation in the stable states leads to a sampling around the minima, but not the barriers. Additional TIS simulations will enhance the sampling of the transition between the minima.

\begin{figure}
    \centering
    \includegraphics[width=1\linewidth]{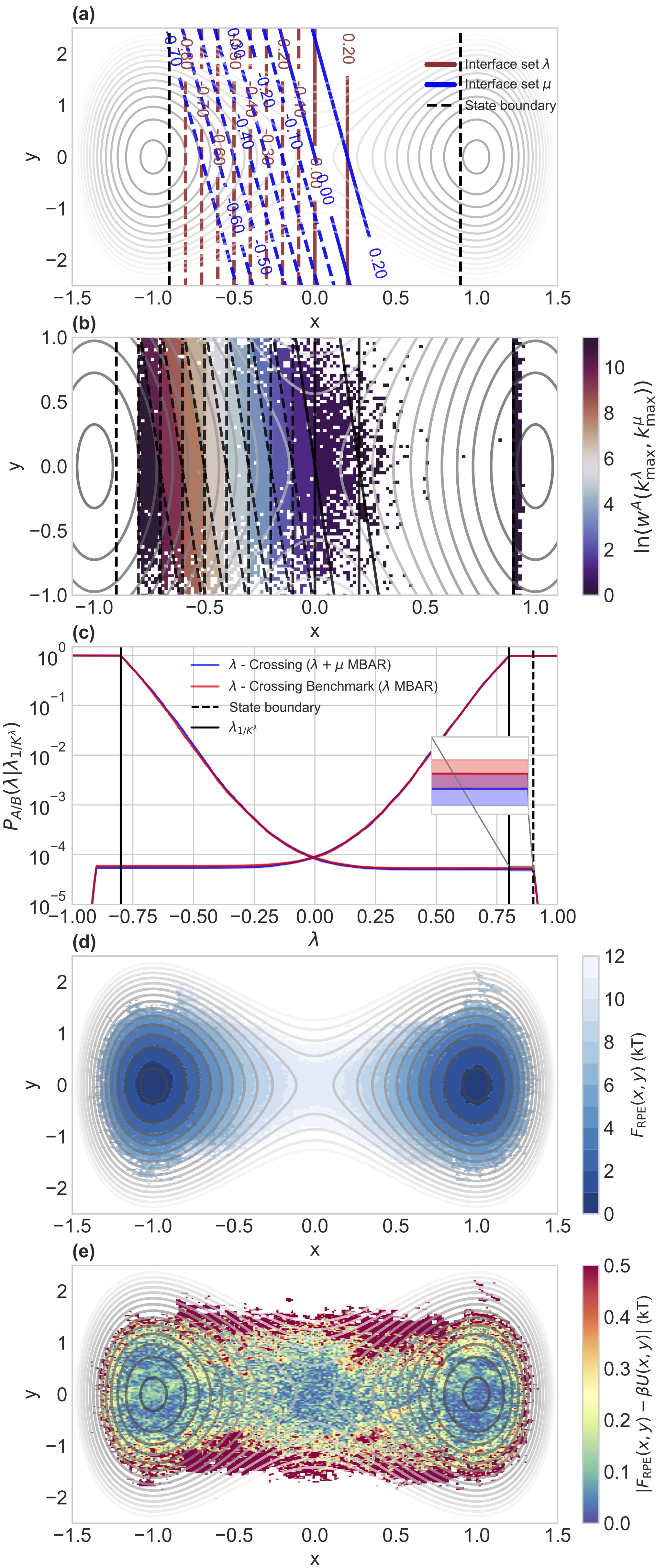}
    \caption{(a) The double well potential with illustration of the two sets of interfaces along $\lambda$ and $\mu$ for the forward ensembles. (b) The weights assigned to each trajectory based on its maximum $k^{max}_{\lambda}$ and $k_{\mu}^{max}$ in the space of $\lambda$ and $\mu$ interfaces. (c) The crossing probability for the forward and backward ensembles computed through two-set MBAR and through a single set benchmark calculation, showing that the two-set procedure recovers the same crossing probability. (d) The free energy of the RPE overlayed with the potential surface. (e) The free energy error relative to the potential energy surface. }
    \label{fig:cross_two_set}
\end{figure}
To investigate the two and MultiSet-MBAR procedure we define a simple interface function with parameters that we can tune to construct multiple different types of interface functions. In particular, we want to investigate a planar interface set that can be rotated in the $x$-$y$ plane under an angle $\theta$. In addition, we would like to be able to add some perturbation towards the linearity, e.g. by invoking some undulation of the planes. To achieve both we define the following function  for sinusoidally perturbed planes,
\begin{equation}
    \lambda(x,y) = x \cos\theta+ y \sin\theta +A_{\rm mag} \sin(2\pi f y),
    \label{eq:interfaceFunctions}
\end{equation}
where $\theta$ is the rotation angle, $f$ the sinusoidal frequency, and $A_{\rm mag}$ the perturbation magnitude.
By varying these parameters, different rotated and modulated interface shapes were generated to assess the robustness of the two and MultiSet-MBAR methods.
To ensure that TIS interfaces do not intersect with the stable state definitions, in OPS %(since the state definition may differ from the interface function) 
the state definition must be explicitly included in the interface ensemble definitions, so that trajectories are only accepted when at least one frame is outside the stable state region, independent of the specific interface function used\cite{swensonOpenPathSamplingPythonFramework2019}.

\subsection{Two-set TIS MBAR results}
As a first test we construct 
the RPE using two interface sets on which TIS is performed.  
The first set  labeled $\{\lambda\}$ consists of interfaces defined perpendicular along the x-axis  encoded by ($\theta=0$, $f=0$ and $A_{\rm mag}=0$). The second set labeled  
$\{\mu\}$ is defined using tilted interfaces encoded by ($\theta=5\degree, f=0, A_{\rm mag}=0$). 
The interface values for these sets $\{\lambda\}=[-0.8, -0.7, -0.6, -0.5, -0.4, -0.3, -0.2, -0.1, 0.0, 0.2]$ and $\{\mu\}=[-0.7, -0.6, -0.5, -0.4, -0.3,  -0.2, -0.1, 0.0,$ $ 0.2]$. 
Fig.~\ref{fig:cross_two_set}a shows an illustration of the two sets.

 We performed TIS employing two-way shooting for each of the interface sets. Shooting points were selected with a gaussian selector with a width of $0.5$ around the interface. The number of trials per interface was $N=5000$. Path lengths were varying from 10 frames in the low interface to 900 frames in the highest interface. Using the maximum reached of each sampled trajectory in both the $\lambda$ and $\mu$ space and solving Eq.~\ref{eq:MBARtwoSetWeight} by iteratively recomputing the partition sums Eq.~\ref{eq:paritionterms_two_set} for each interface, we obtain the weight for each trajectory. The same procedure can also be used for the backward ensembles but now using the minimum reached.

Fig.~\ref{fig:cross_two_set}b plots the weights assigned based on where in the $x$-$y$ plane the path reaches its maximum, highlighting how the weights of the paths are now based on the maxima $k_{max}^{\lambda}$ and $k^{\mu}_{max}$ in the space of $\lambda$ and $\mu$ interfaces.

Using Eq.~\ref{eq:crosprob_twoSet}, the crossing probability for the forward and backward ensembles using both sets is computed along the $\lambda$ function as shown in Fig.~\ref{fig:cross_two_set}c. A bootstrap procedure is applied on the dataset of 4900 paths per interface (coming from a 5000 MC step TIS simulation for which the first %$N_{thermalization}=$
$100$ paths have been removed). For the bootstrap procedure a blocksize of 10 paths is used to reflect the correlation in the dataset and 100 bootstrap resampling are performed, giving a statistical error on the crossing probability and a mean as shown in the plot. The crossing probability plots in Fig.~\ref{fig:cross_two_set}c show  that the two-set procedure recovers within the statistical error  the same crossing probability $P_A(\lambda|\lambda_1)$ for $\lambda(x)=x$ as the single set procedure for a benchmark TIS calculation on the $\lambda$ interface set consisting of $N=10000$ paths per interface.

To evaluate the accuracy of the RPE free energy relative to the true free energy surface, we use the mean absolute error (MAE). However, since some bins at the edges contain very few points, they can disproportionately affect the overall error. To mitigate this, we compute a weighted MAE where each bin is weighted by its local point density $h(x,y)$, ensuring that sparsely sampled regions contribute less. Figure~\ref{fig:cross_two_set}d shows the RPE Free energy projected onto the $x\text{-}y$ plane, where states are aligned by the flux through the shared first interface ($\lambda_1 \cap \mu_1$). The deviation from the true potential $\beta U_{\rm dw}(x,y)$ is shown in Fig.~\ref{fig:cross_two_set}e. Peripheral regions show higher errors due to limited sampling. To account for this we define the weighted MAE:
\begin{align}
    \text{MAE} = \sum_{x,y} h(x,y)\left|F_{\text{RPE}}(x,y) - \beta U_{dw}(x,y)\right|,
    \label{eq:weighted_MAE}
\end{align}
where \( h(x,y) \) is the normalized histogram of RPE points per bin. We use the weighted MAE below to evaluate the quality of the RPE.

\subsection{Multiple-Set TIS MBAR}

Above we showed how the two-set MBAR procedure can be used to combine two sets of TIS interface ensembles sampled according to different interface functions $\lambda,\mu$. This procedure  generalizes to $M$ sets where the weights are determined by the maximum reached in the joined space of interface functions Eq.\ref{eq:multiset_MBAR_weight}. 
The MultiSet-MBAR procedure ensures consistent relative scaling of path weights across different ensembles, as the unbiased weights are obtained from partition functions that couple all sampled path ensembles and explicitly account for their mutual overlap.

\begin{figure*}
    \centering
    \includegraphics[width=1\linewidth]{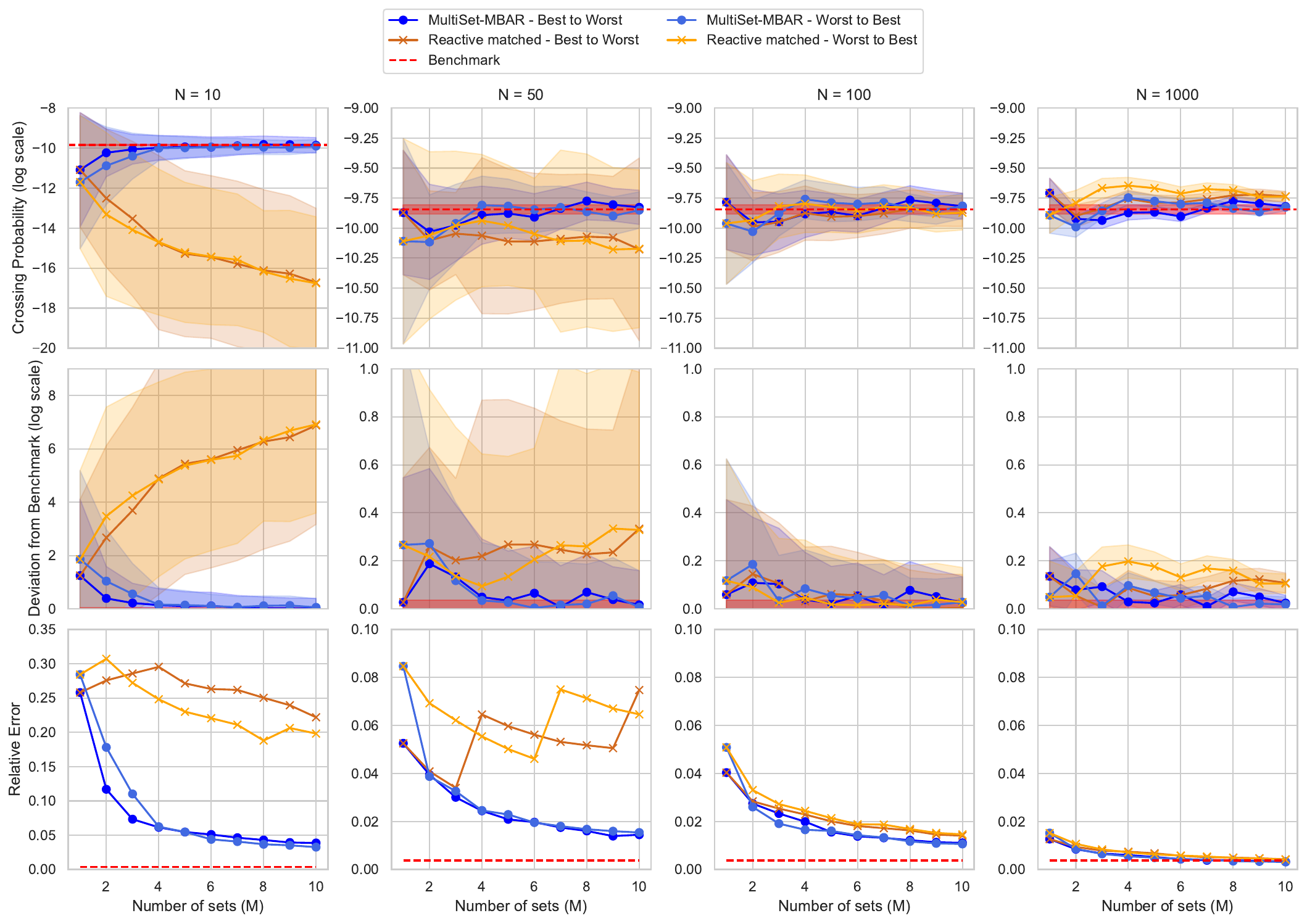}
      \caption{ (top row)
    Estimated crossing probability $P_A(\lambda_B|\lambda_1)$ using $\lambda(x) = x$ as a function of the number of sets $M$ being combined, for different number of paths per interface, $N$, for both the reactive matching method and the MultiSet-MBAR approach. Results are compared to a benchmark computed from $N=10000$ paths sampled along the $x$ interface set. The MultiSet-MBAR estimate converges smoothly to the benchmark as more sets are added, while the independent reactive matching estimate initially diverges for small $N$ and only aligns at larger sample sizes.
    (middle row) Logarithmic deviation of the estimated crossing probabilities from the benchmark value as a function of the number of sets $M$ being combined, for different number of paths per interface, $N$, for both the independent reactive matching method and the MultiSet-MBAR approach. The MultiSet-MBAR method consistently reduces deviation as more data is included, demonstrating improved convergence. In contrast, the reactive matching approach shows increased deviation for low $N$, especially as additional sets are added.
    (bottom row) Relative statistical error of the crossing probability estimates, defined as $\sigma_{\text{bs}}/\mathbb{E}[\log P_A(\lambda_B|\lambda_1)]_{\text{bs}}$, as a function of the number of sets $M$ being combined, for different number of paths per interface, $N$, for both the independent reactive matching method and the MultiSet-MBAR approach. The MultiSet-MBAR method exhibits a consistent reduction in relative error, approximately following a $1/\sqrt{M}$ trend with increasing number of sets. In contrast, the reactive matching  approach shows a plateau in error for small $N$, only improving when the sample size becomes sufficiently large.
    }
    \label{fig:cross_results}
\end{figure*}

\begin{figure*}
    \centering
    \includegraphics[width=1\linewidth]{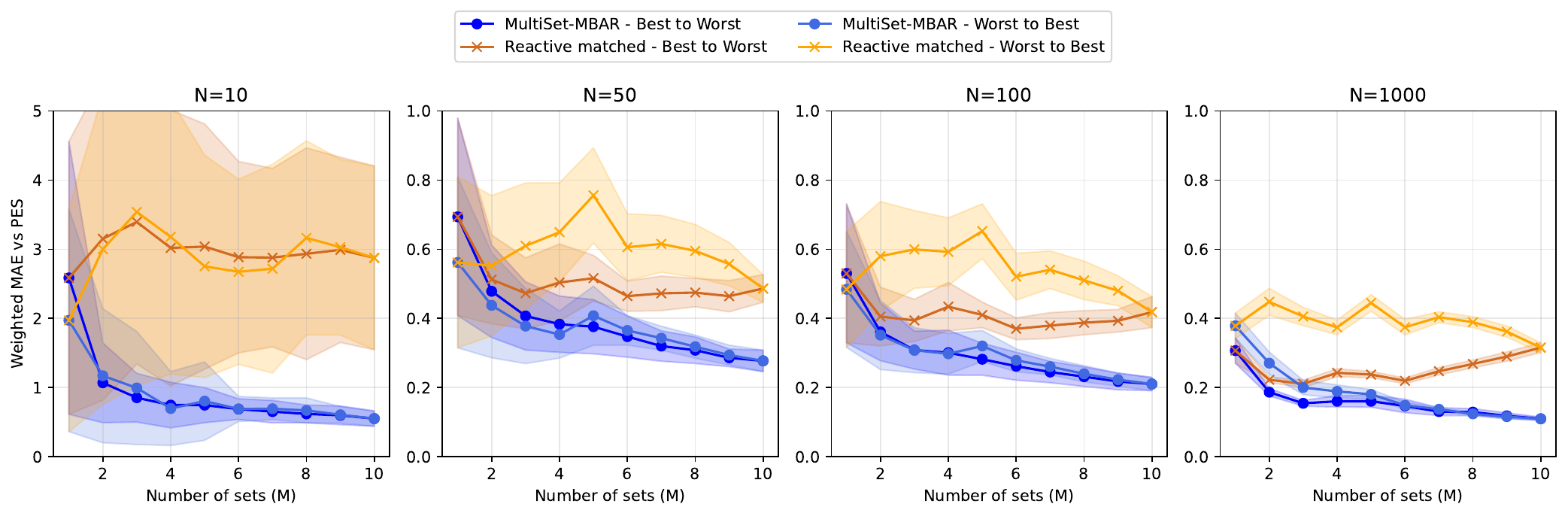}
    \caption{Weighted mean absolute error (MAE) between RPE free energy and potential energy surface, weighted by the density of trajectory configurations per bin, as the number of sets increases for N=10, 50,100,1000, using bootstrapping with 100 repeated resampling of $N$ trajectories to get a statistical error.}
    \label{fig:MAE_RPE_vs_M}
\end{figure*}

\subsubsection{Crossing probability analysis}

First, we focus on the effect of the reweighting procedure in the crossing probability
\begin{equation}
    P_A(\lambda_B|\lambda_1) = \sum^{RPE}_{\XP\in\{\XP\}^{(M)}} w[\XP]\theta(\lambda^{\text{max}}[\XP]-\lambda)h_{A,\lambda_1}[\XP],
\end{equation}
computed as in Eq.~\ref{eq:crosproball}, but now summing over all weights $\{w\}$ and paths $\{\XP\}$ in the combined RPE of $M$ sets of simulations. Here we use $\sum^{RPE}_{\XP\in\{\XP\}} = \sum^{M}_{i}\sum^{K^{(i)}}_{k}\sum^{N^{(i)}_{k}}_{n}$ for clarity.
To evaluate the improvement offered by the MultiSet-MBAR reweighting over a selected rescaling combination of independent RPEs, we construct the sets $\{w\}^{(M)}$ and $\{\XP\}^{(M)}$ using two different approaches.

In the MultiSet-MBAR approach, weights are assigned according to Eq.~\ref{eq:multiset_MBAR_weight} for each path $\XP\in\{\XP\}^{(M)}\equiv\{\{\XP\}^1,\cdots\{\XP\}^{M}\}$, where $\{\XP\}^{(M)}$ is the merged set of paths from $M$ separate TIS simulations, each with a different set of interfaces. 

In an alternative rescaling approach (referred to as reactive matched), weights are computed independently for each subset $\{\XP\}^i$ using Eq.\ref{eq:SingleSetMBARWeights}. Because each individual weighting carries an arbitrary normalization constant, we rescale all weights so that the weight of reactive paths is unity. 
All paths from each subset are then combined into one RPE. As another option, one could also rescale the weights by matching the weight through the first interface of each set to the flux determined through an independent MD simulation in the stable basins. In the two alternative rescaling approaches partition terms  $\ZZ^{k}_{A\lambda^{(i)}}$ for each interface ensembles set $\lambda^{(i)}$ are determined independently, each with its own (unknown) relative offset. 
In contrast, the MultiSet-MBAR method naturally aligns these partition terms on a common scale by incorporating all ensemble distributions simultaneously during reweighting.

We compare the MultiSet-MBAR and  reactive matched approaches across $M=10$ interface functions (see Fig.\ref{fig:SI:intefaces_10_sets} in the Appendix for an illustration).
For each interface function $\lambda^{(i)}$ we perform 10 TIS simulations, covering the range $\lambda_1^{(i)}=-0.8$ to $0.0$ in increments of $0.1$ and an additional interface at $0.2$. Thus each interface set $i\in\{1,\ldots,M\}$ contributes ensembles $P^k_{A,\lambda^{(i)}}$ for interface indices $k\in\{1,\ldots,K^{(i)}\}$, sampled with TIS conditional paths on $\lambda^{(i)}$.
As the  order of adding new interface sets to the weighting influences the final result, we will compare two different orders, labeled by 'best to worst' and 'worst to best'. The distinction between the two orders is based on how much the interface function differs from the ideal set of interfaces (in this case the $x$-interfaces), using a simple scoring function (see Eq.~\ref{eq:interface_score} in the Appendix).  

In the bootstrap procedure, we draw $N$ paths from each TIS ensemble $P^{k}_{A,\lambda^{(i)}}$, with a block-size of 1, producing a largely uncorrelated set of paths per ensemble and repeat this $N_{\rm bs}=100$ times to get a statistical error. 

In Fig.~\ref{fig:cross_results}a, we show how the predicted crossing probability $ \log p_A(\lambda_B|\lambda_1)$ changes with the number of sampled paths per interface, $N=10,50,100$ and $1000$. The deviation of these results relative to a benchmark obtained using $N=10000$ paths sampled along the $x$-interface set is shown in Fig.~\ref{fig:cross_results}b.

As more interface sets are included, the MultiSet-MBAR estimate converges to the benchmark value. In contrast, the reactive matched approach diverges from the benchmark for small $N$, with its accuracy degrading as more sets are added. As $N$ increases, both methods begin to agree more closely, the reactive matched method also starting to closely match the benchmark for $N\ge 100$.

In Fig.~\ref{fig:cross_results}c, we present the relative error, defined as $\sigma_{\text{bs}}/\mathbb{E}(\log P(\lambda_B|\lambda_1))_{\text{bs}}$, as a function of the number of sets $M$. For small $N$, the reactive matched  error remains roughly constant or plateaus, even as more sets are added. In contrast, the MultiSet-MBAR error consistently decreases, approximately following a $1/\sqrt{M}$ scaling, where $M$ is the number of sets. When $N$ becomes sufficiently large, the reactive matching based approach also shows decreasing error, but remains less accurate than MultiSet-MBAR.
 
These results highlight that for small sample size, the MultiSet-MBAR approach consistently improves as more data is incorporated, whereas the reactive based independent rescaling method can degrade. With sufficient data, both methods benefit from additional sets, but MultiSet-MBAR maintains a lower statistical error throughout.

Note that the order of adding interfaces does apparently have relatively little influence on the   accuracy of the results, since both the two orange and the two blue curves are close to each other. However, when looking more carefully we see that for just a single set (M=1), the error is larger for the worst-to-best order,  as expected, although the error quickly drops and then (naturally) converges to the same answer.
The reason for this relative insensitivity  to the order of the interface set choice is probably that all interface sets are capable of sampling the transitions. Indeed, this is one of the hallmarks of TIS\cite{vanErp2003}.

\subsubsection{RPE distribution and Free energy}

By constructing the RPE using the MultiSet-MBAR approach, we also obtain an estimate of the configurational distribution and free energy of the system. Figure~\ref{fig:cross_two_set}b shows a projection of the resulting free energy, which includes stable state data matched via the procedure described in Sec.~\ref{sec:RPEconstruction}.

To quantify how the inclusion of additional sets improves the RPE, we compute the weighted mean absolute error using Eq.~\ref{eq:weighted_MAE} between the MultiSet-MBAR free energy estimate and the potential energy surface. Specifically, we focus on the transition region between the two stable states (defined here as $x > -0.9$ and $x < 0.9$), which is particularly sensitive for both crossing probabilities and free energy estimation.

In Fig.~\ref{fig:MAE_RPE_vs_M}, we show how the weighted MAE decreases as a function of the number of combined sets used in the MultiSet-MBAR reweighting. We assess this trend for different sample sizes $N = 10, 50, 100, 1000$, using a bootstrapping procedure sampling $N$ samples from each ensemble to estimate statistical uncertainties. 

The results indicate that the inclusion of more sets systematically improves the accuracy of the RPE, as reflected by the decreasing MAE. This highlights the robustness of the MultiSet-MBAR method in integrating information from multiple TIS simulations to refine the free energy landscape. In contrast, the reactive matching seems not to converge, even for the largest sample size.

This analysis complements the crossing probability discussion in the previous section: both observables depend on accurate path weight estimation, and their mutual convergence further validates the reliability of the reweighting approach.

\subsection{Host Guest system from AIMMD-TIS calculations}

To illustrate the utility of MultiSet-MBAR beyond simple model systems, we apply it to committor-based TIS on a solvated Host–Guest system from the AIMMD-TIS study \cite{breebaart2025understandingreactionmechanismsstart}. In this iterative framework, an initial committor model (defined as $p_B(x)=1/({1+e^{-q(x|\theta)}})$ with $q(x|\theta)$ the neural net model) learned from TPS shooting data using AIMMD\cite{Jung2023} (indicated as $q(x|\theta_{TPS})$) defines isocommittor interfaces for a first TIS calculation, generating path ensembles that can be reweighted via the RPE. 

The resulting RPE data is then used to train an improved committor model $q(x|\theta_{RPE_1})$, which can in turn define new interfaces for the next TIS iteration - repeating the cycle. Unlike standard approaches where previous path data is discarded due to differing biases in each iteration, MultiSet-MBAR enables the statistically consistent reuse of all TIS path ensembles across iterations. Each ensemble, conditioned on a different committor model, is treated as a separate biased ensemble and jointly reweighted to a common underlying equilibrium path distribution.

\begin{figure}
    \centering
    \includegraphics[width=1\linewidth]{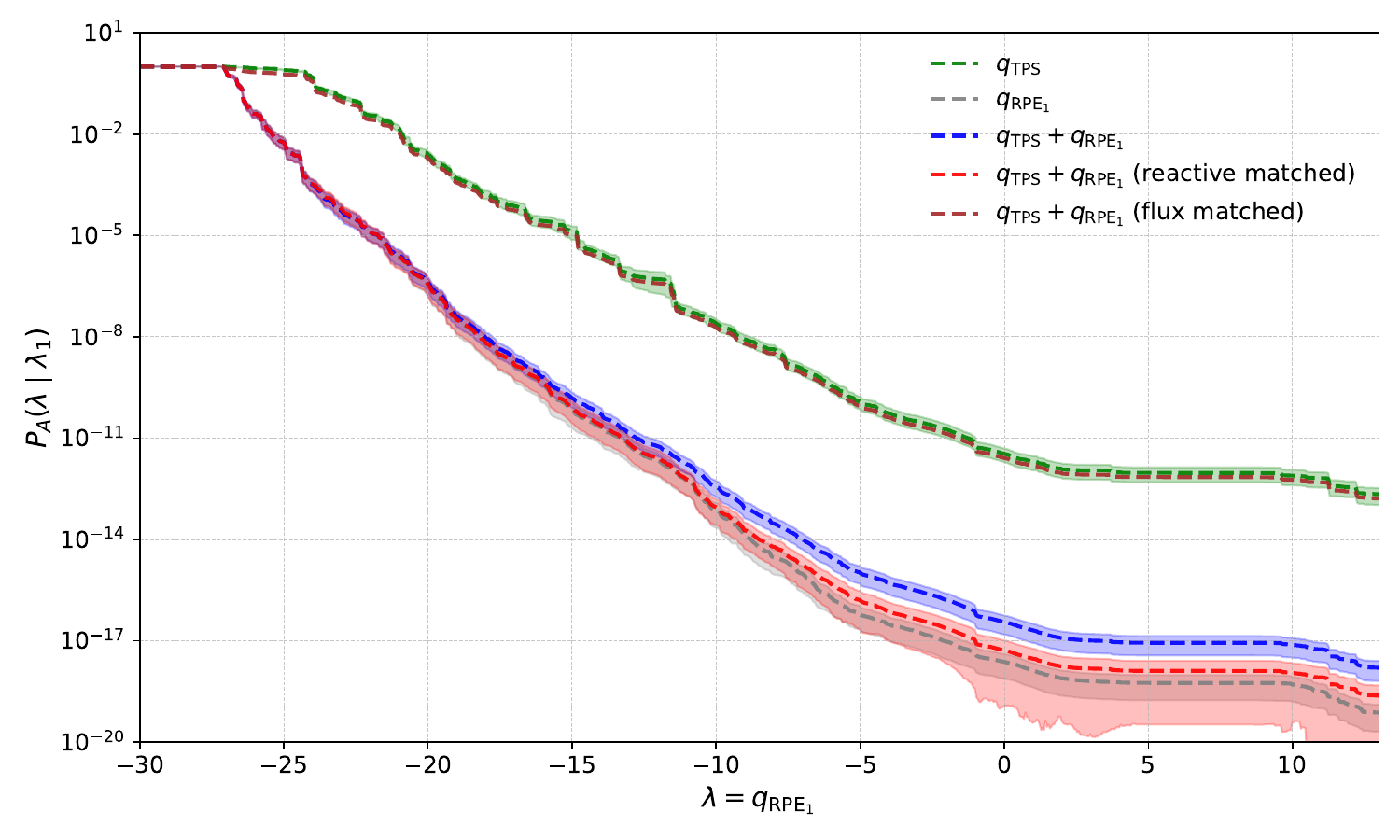}
    \caption{
    Forward crossing probability $P_A(q(x|\theta_{\mathrm{RPE}_1}) \mid q_A)$ evaluated using reweighted path ensembles from successive TIS iterations on a committor model $q(x|\theta)$ for a host-guest system. Independent RPEs constructed from TIS simulations using $q(x|\theta_{\mathrm{TPS}})$ (green) and $q(x|\theta_{\mathrm{RPE}_1})$ (grey) are compared with their joint reweighting using MultiSet-MBAR (blue). Also shown are independent RPE rescaling strategies based on reactive matching (red) and flux matching (brown). Due to the extremely large statistical uncertainty associated with flux matching, its error bars are omitted for clarity,  the relative error is reported in the text.
    }
    \label{fig:Host-Guest_crossing_probability}
\end{figure}

Fig.~\ref{fig:Host-Guest_crossing_probability} shows the forward crossing probability $P_A(q(x|\theta_{\mathrm{RPE}_1}) \mid q_A)$ evaluated using different RPE constructions: an RPE constructed from TIS data generated using $q(x|\theta_{\mathrm{TPS}})$, an RPE constructed from TIS data generated using $q(x|\theta_{\mathrm{RPE}_1})$, and a combined RPE obtained by jointly reweighting both datasets using the MultiSet-MBAR approach.
In addition, we compare two independent rescaling strategies in which RPEs are combined after reweighting. In the first, referred to as reactive matching,  and also shown in the comparison for the double well results, the MBAR weights are rescaled such that all reactive paths have unit weight. In the second, referred to as flux matching, the weights are rescaled to match the flux through the first interface to an independently computed molecular dynamics flux through that interface. 
Here, $q_A$ is defined as the maximum value of $q(x|\theta_{\mathrm{RPE}_1})$ sampled in a molecular dynamics simulation constrained to state $A$, providing a practical boundary beyond which configurations are considered outside the state.

The results clearly show that the first iteration alone is insufficient, whereas the second iteration leads to a substantial improvement in the estimated crossing probability. Combining both datasets using MultiSet-MBAR further improves the statistical quality by jointly reweighting all available path ensembles. In contrast, independent rescaling strategies depend strongly on the chosen matching criterion. Reactive matching yields a curve closer to that obtained from the lower-crossing-probability RPE, as the corresponding weights dominate the combined estimate. Flux matching yields a mean crossing probability much closer to that obtained from the higher-crossing-probability RPE, since this dataset dominates after rescaling.
While the mean crossing probability obtained from flux matching is shown in Fig.~\ref{fig:Host-Guest_crossing_probability}, we omit its  statistical uncertainty 
as it is extremely large due to the orders-of-magnitude variation in the rescaled weights.

Using a bootstrap analysis with block size $10$, we find relative errors in the logarithm of the mean crossing probability of $1.68\%$ for the $q(x|\theta_{\mathrm{TPS}})$ dataset, $1.66\%$ for $q(x|\theta_{\mathrm{RPE}_1})$, $2.36\%$ for the reactive matched combination, $14.18\%$ for the flux-matched combination, and $1.47\%$ for the MultiSet-MBAR combined RPE. These results demonstrate that MultiSet-MBAR yields the lowest statistical uncertainty while avoiding the dominance bias introduced by independent rescaling strategies.

\section{Conclusions}
We developed the MultiSet-MBAR  method to combine TIS simulations based on different CV functions. This approach enables the reuse of trajectories generated with different interface definitions in a unified framework, resulting in a more accurate and representative Reweighted Path Ensemble that incorporates trajectories from multiple sets. 
The  MultiSet-MBAR approach is significantly more accurate than an independent rescaled combination for (reweighted) path ensemble. 
In contrast to the simple rescaling approaches, which effectively assigns independent partition sums  for each interface ensembles, each with its own (unknown) relative offset, the MultiSet-MBAR method naturally aligns these partition sums on a common scale by incorporating all ensemble distributions simultaneously during reweighting.

Our novel method can facilitate the iterative optimization of the interface functions used in TIS, even those based on neural net  committor models,  while preserving the ability to reuse previously sampled trajectories. This allows for an efficient, feedback-driven procedure in which the choice of interface function can be refined over successive TIS iterations, leading to improved sampling and more reliable mechanistic insight.

\section*{Data Availability Statement}
The data that supports the findings of this study are available within the article. The code and scripts to produce and analyze the data are available online.

\appendix
\section{Derivations of MBAR equations}
\subsection{Consistency check of Eq.~\ref{eq:log_max_lambda_weight} and Eq.~\ref{eq:log_max_mu_weight}}
Here we check the consistency of the two-set MBAR procedure with the standard single set MBAR.
Starting from  Eq.\ref{eq:log_max_lambda_weight} and Eq.\ref{eq:log_max_mu_weight},  
if we assume $\mu=\lambda$ and the set of interfaces $K_1=K_2\equiv K$
then this should give the same result as MBAR with a single interface set but now running over $N_k+\tilde N_k$ paths per $k$'th simulation:
\begin{align}
    \frac{\partial \ln L}{\partial  w_{j,n} } = 0 = \frac{1}{w_{j,n}} - \sum_{k}^{K}  \frac{(N_k+\tilde N_k)  \hat{h}_{A,\lambda}^k[\XP_{j,n}]} {\mathcal{Z}_{A,\lambda}^{k}}
\end{align}
\begin{align}
    \frac{\partial \ln L}{\partial  \tilde{w}_{j,n} } = 0 = \frac{1}{\tilde{w}_{j,n}} -  \sum_{k}^{K}  \frac{(N_k+\tilde N_k)  \hat{h}_{A,\lambda}^k[\tilde{\XP}_{j,n}]} {\mathcal{Z}_{A,\lambda}^{k}}, 
\end{align}
which is the same as the original single set MBAR log likelihood expression Eq.~\ref{eq:singleSetLoglikelihood} but now with $N_k+\tilde N_k$ paths.
This show that the two set MBAR procedure leads to consistent results in the limit of identical sets. 

\begin{table}[b]
\caption{Notation used in the MBAR/RPE derivations.}
\begin{tabular}{p{0.23\linewidth} p{0.70\linewidth}}
\hline
Symbol & Meaning \\
\hline
$\XP$                                   & Trajectory (path) in path space \\
$\PP[\XP]$                              & Unbiased path distribution \\
$h^{k}_{A,\lambda^{(i)}}[\XP]$           & Indicator: $\XP$ crosses interface $k$ of set $i$ leaving $A$ \\
$\lambda^{(i)}$                         & Interface (CV) function for set $i$, with interfaces $\{\lambda^{(i)}_k\}_{k=0}^{K^{(i)}}$ \\
$K^{(i)}$                               & Number of interfaces in set $i$ \\
$N^{(i)}_k$                             & Number of sampled paths in interface ensemble $k$ of set $i$ \\
$\ZZ$                                   & Global normalization constant  \\
$\ZZ^{k}_{A,\lambda^{(i)}}$             & Conditional partition sum for interface $k$ in set $i$ (leaving $A$) \\
$w^{A}[\XP]$                            & MBAR weight assigned to path $\XP$ leaving state $A$\\
$K^{(i)}_{\max}[\XP]$                   & Highest interface crossed by $\XP$ in set $i$ \\
$P_A(\lambda\mid \lambda_1)$            & Crossing probability conditioned on reference interface $\lambda_1$ \\
$c_A,c_B$                               & Constants to match forward/backward reactive histograms and flux \\
\hline
\end{tabular}
\end{table}

\subsection{Derivation MultiSet-MBAR}
\label{A:MultiSetMBAR_Derivation}

In this section we provide the generalization of the derivation provided in  Sec.~\ref{sec:twosets} for two sets to $M$ interface ensemble sets. Consider  $M$ interface functions $\lambda^{(1)},\ldots,\lambda^{(M)}$ for $i\in1,\ldots,M$, which each have a $N^{(i)}_{k}$ for $k\in 1,\ldots K^{(i)}$ where $K^{(i)}$ is the number of interfaces in set $i$ and $N^{(i)}_{k}$ indicates the number of paths in interface ensemble $k$ (in $\lambda^{(i)}$). Each set $i$ also has a corresponding set of conditional partition functions $\ZZ_{A,\lambda^{(i)}}^{k}$ for paths leaving $A$ and $\ZZ_{B,\lambda^{(i)}}^{k}$ for paths leaving B.

We assume that for each set the stable states are identical ($\lambda_A\equiv \lambda^{(i)}_{0}$ for all $i\in 1,\ldots,M$).
For each interface ensemble $i,k$ we obtain a set $\{\XP_n\}^{(i)}_{k}$ containing $N^{(i)}_{k}$ paths where $\XP^{(i)}_{k,n}$ is the $n$'th path in ensemble $k$ of interface function $\lambda^{(i)}$.

We can join the $M$ sets to get the probability that we observe them in an unbiased ensemble as
\begin{align}
    \PP_{A}[\XP] = \frac{1}{\ZZ} \left( \sum^{M}_{i}\sum_{k}^{K^{(i)}} \sum_n^{N_k^{(i)}}  w^{(i)}_{k,n} \delta[\XP-\XP^{(i)}_{k,n}] \right),
    \label{eq:multisetUnbiased}
\end{align}
where the factor 
\begin{equation}
    \ZZ_{A} = \sum^{M}_{i}\sum_{k}^{K^{(i)}} \sum_n^{N_k^{(i)}} w^{(i)}_{k,n},
\end{equation}
with unknown weights $w^{(i)}_{k,n}$.

If we consider the  (TIS) ensemble from which we obtain the conditional distribution first in a theoretical sense, where we can sample all paths from the equilibrium path distribution $\PP[\XP]$, then 
\begin{align}
    \PP^{k'}_{A,\lambda^{(i)}}[\XP] =& \frac{1}{Z^{k'}_{A,\lambda^{(i)}}} \int \DD \XP' \hat{h}_{A,\lambda^{(i)}}^{k'}[\XP']\PP[\XP']\delta[\XP-\XP']
\end{align}
where $Z^{k'}_{A,\lambda^{(i)}}= \int \DD \XP' \hat{h}_{A,\lambda^{(i)}}^{k'}[\XP']\PP[\XP']$ is the theoretical conditional partition function.

In the discrete case the distribution of paths which satisfy the bias function  $\hat{h}_{A\lambda^{(i)}}^k[\XP]$ is given by:
\begin{align}
    \PP^{k'}_{A,\lambda^{(m)}}[\XP] =& \frac{1}{\mathcal{Z}^{k'}_{A,\lambda^{(m)}}}\sum^{M}_{i}\sum_{k}^{K^{(i)}} \sum_n^{N_k^{(i)}} \hat{h}_{A,\lambda^{(m)}}^{k'}[\XP]\PP[\XP]\delta[\XP-\XP_{k,n}] \\
 =& \frac{1}{\mathcal{Z}^{k'}_{A,\lambda^{(m)}}}\sum^{M}_{i}\sum_{k}^{K^{(i)}} \sum_n^{N_k^{(i)}} \hat{h}_{A,\lambda^{(m)}}^{k'}[\XP] w^{(i)}_{k,n} \delta[\XP-\XP^{(i)}_{k,n}],
\end{align}
with $\mathcal{Z}^{k'}_{A,\lambda^{(m)}}=\sum^{M}_{i}\sum_{k}^{K^{(i)}} \sum_n^{N_k^{(i)}} \hat{h}_{A,\lambda^{(m)}}^{k'}[\XP] w^{(i)}_{k,n}$.

The paths $\{\XP\}^{(i)}_{k}$ are sampled from distribution $\PP_{A\lambda^{(i)}}^{k}[\XP]$, such that the likelihood of the path $\XP^{(i)}_{k,n}$ which is drawn from the $k$'th interface ensemble with interface function $\lambda^{(i)}$ is given by 
\begin{equation}
    L_{k,\lambda^{(i)}}(\XP^{(i)}_{k,n}|\{w_{k,n}\}^{1},\ldots,\{w_{k,n}\}^{M}) = \PP_{A\lambda^{(i)}}^{k}[\XP^{(i)}_{k,n}].
\end{equation}
The log likelihood of the observed data of the $M$ sets is thus 
\begin{align}
    \ln L =& \ln \prod^{M}_{i} \prod_{k}^{K^{(i)}} \prod_{n}^{N_k^{(i)}} \PP_{A,\lambda^{(i)}}^k[\XP^{(i)}_{k,n}]  \notag \\
    =& \ \sum^{M}_{i} \sum_{k}^{K^{(i)}} \sum_{n}^{N_k^{(i)}} \ln\PP_{A,\lambda^{(i)}}^k[\XP^{(i)}_{k,n}].
\end{align}
substitution of the distribution gives 
\begin{align}
    \ln L = \sum^{M}_{i} \sum_{k}^{K^{(i)}} \sum_{n}^{N_k^{(i)}} ( \ln  w^{(i)}_{k,n} \hat{h}_{A,\lambda^{(i)}}^{k}[\XP_{k,n}]  - \ln  {Z}^{k}_{A,\lambda^{(i)}}).
\end{align}
Maximizing this w.r.t. $w^{(m)}_{j,n}$ 
\begin{align}
    \frac{\partial \ln L}{\partial  w^{(m)}_{j,n} } = 0 = \frac{1}{w^{(m)}_{j,n}} - \sum^{M}_{i=1}\sum_{k}^{K^{(i)}}  \frac{N^{(i)}_k  \hat{h}_{A,\lambda^{(i)}}^k[\XP^{(m)}_{j,n}]} {\mathcal{Z}_{A,\lambda^{(i)}}^{k}}
    \label{eq:log_max_lambda_weight_multiset}
\end{align}
\begin{align}
   \frac{w^{(m)}_{j,n}}{\ZZ} = \left[\sum^{M}_{i=1}\sum_{k}^{K^{(i)}}  \frac{N^{(i)}_k  \hat{h}_{A,\lambda^{(i)}}^k[\XP^{(m)}_{j,n}]} {\mathcal{Z}_{A,\lambda^{(i)}}^{k}/\ZZ}\right]^{-1}
    \label{eq:log_max_lambda_weight_multiset}
\end{align}
\begin{equation}
    \frac{\mathcal{Z}_{A,\lambda^{(i)}}^{k'}}{\ZZ} = \sum^{M}_{i=1}\sum_{j}^{K^{(i)}}\sum_n^{N^{(i)}_j}  \hat{h}_{A,\lambda^{(i)}}^{k'}[\XP^{(i)}_{j,n}]\left[\sum^{M}_{i=1}\sum_{k}^{K^{(i)}}  \frac{N^{(i)}_k  \hat{h}_{A,\lambda^{(i)}}^k[\XP^{(i)}_{j,n}]} {\ZZ_{A,\lambda^{(i)}}^{k}/\ZZ}\right]^{-1}, 
\end{equation}
which can be solved iteratively solving $\sum_{i=1}^{M}K^{(i)}$ partition terms. 
A path $\XP$ gets a weight

\begin{equation}
    w^{A}[\XP]=\left[\sum^{M}_{i=1}\sum_{k}^{K^{(i)}}  \frac{N^{(i)}_k  \hat{h}_{A,\lambda^{(i)}}^k[\XP]} {\ZZ_{A,\lambda^{(i)}}^{k}/\ZZ}\right]^{-1}, 
\end{equation}
which is only based on the maximum of the path $\XP$ in each of the $M$ monotonically increasing $\lambda$ functions (as all latter contributions in the sum have $h^{k}_{A\lambda^{(i)}}=0$ for $k>K_{\rm max}^{(i)}[\XP]$). So 
\begin{equation}
    w^{A}[\XP]=\left[\sum^{M}_{i=1}\sum_{k}^{K_{\rm max}^{(i)}[\XP]}  \frac{N^{(i)}_k} {\ZZ_{A,\lambda^{(i)}}^{k}/\ZZ}\right]^{-1}, 
\end{equation}
for all paths, independent from the distribution they where originally sampled from.

\section{Interface Functions and Deviation Scores}
\label{A:interface_functions_score}
\begin{figure}
    \includegraphics[width=1\linewidth]{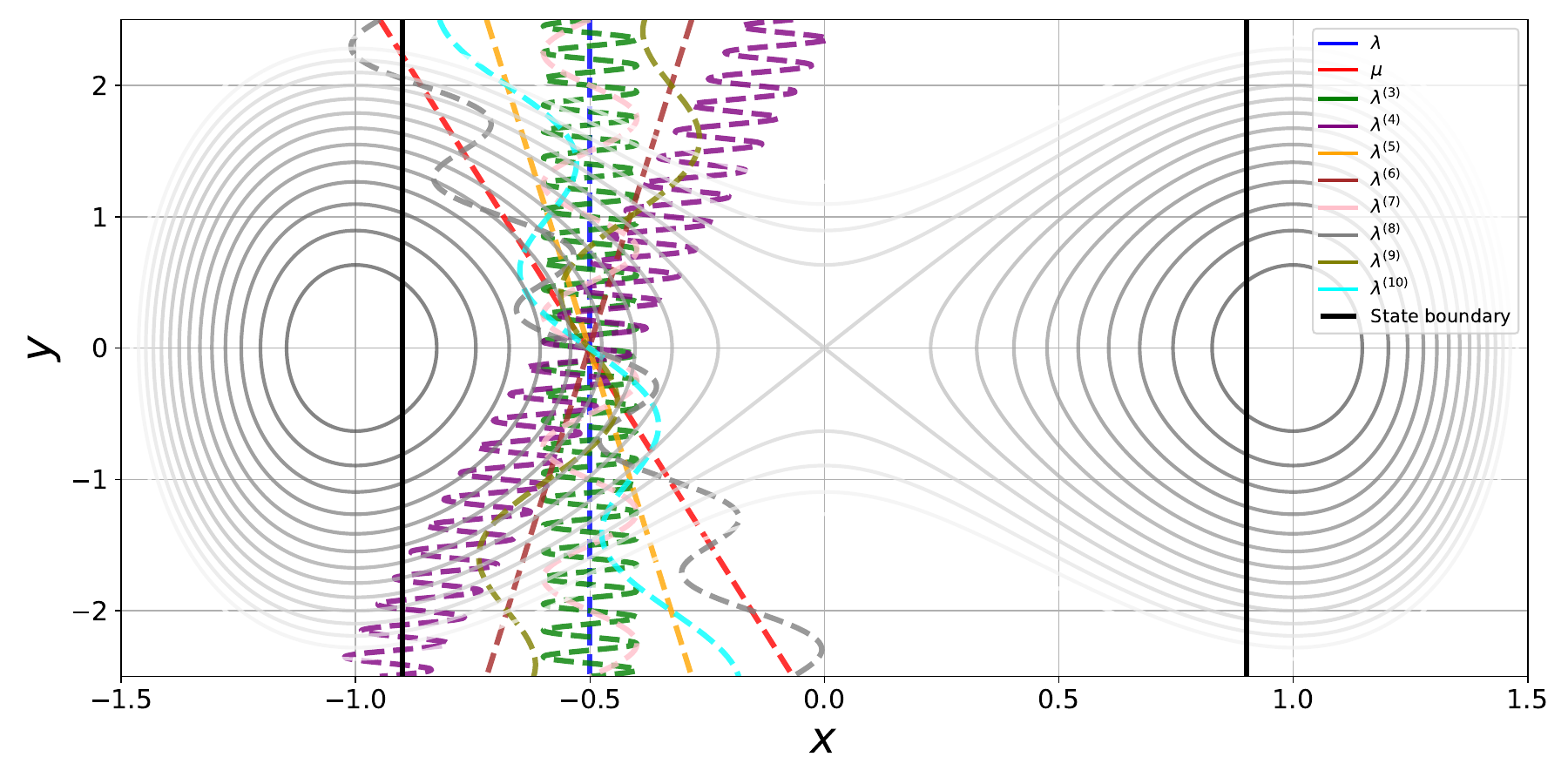}
    \caption{Double well potential with an illustration of an interface for each of the 10 different interface functions.}
    \label{fig:SI:intefaces_10_sets}
\end{figure}
To test the MultiSet-MBAR procedure we define a family of interface functions $\lambda(x, y)$ with modulated the shape of the interfaces in space, parameterized by a rotation angle $\theta$, a sinusoidal frequency $f$, and a magnitude of oscillation $A_\text{mag}$. The general form is given in Eq.~\ref{eq:interfaceFunctions}. We evaluate how much each interface deviates from a flat interface aligned with the $x$-axis with a score function:
\begin{equation}
\text{score}(\theta, f, A_{\text{mag}}) = 
\begin{cases}
|\theta| + b \cdot A_{\text{mag}}, & \text{if } f = 0 \\
|\theta| + a \cdot \left(\frac{1}{f}\right) + b \cdot A_{\text{mag}}, & \text{if } f \neq 0
\end{cases}
\label{eq:interface_score}
\end{equation}
which penalizes rotation and with $a = 2$ and $b = 100$ as weighting parameters that penalize low frequencies and large oscillation magnitudes, respectively.

Table~\ref{tab:interface-functions} lists the ten considered interface functions, their parameter values and computed scores. Functions are ordered by increasing deviation score. The first two are denoted $\lambda$ and $\mu$ as they have been used for the two-set procedure, and the remaining are labeled $\lambda^{(i)}$ for $i \geq 3$.  In Fig.~\ref{fig:SI:intefaces_10_sets}, a single interface of each of the 10 different interface functions is illustrated.

\begin{table}[h!]
\centering

\begin{tabular}{ccccc}
\toprule
Symbol & $\theta$ [deg] & $f$ [Hz] & $A_{\text{mag}}$ & Score \\
\hline
$\lambda$       & 0.0   & 0.0 & 0.0 & 0.00 \\
$\mu$           &  5.0   & 0.0 & 0.0 & 5.00\\
$\lambda^{(3)}$   &  -5.0  & 0.0 & 0.0 & 5.00\\
$\lambda^{(4)}$   &  10.0  & 0.0 & 0.0 & 10.00\\
$\lambda^{(5)}$   & 0.0 & 5.0 & 0.1 & 10.40 \\
$\lambda^{(6)}$   & 0.0 & 1.0 & 0.1 & 12.00 \\
$\lambda^{(7)}$   &  -5.0 & 0.5 & 0.1 & 19.00 \\
$\lambda^{(8)}$   &   5.0 & 0.5 & 0.1 & 19.00 \\
$\lambda^{(9)}$   &  -10.0 & 5.0 & 0.1 & 20.40 \\
$\lambda^{(10)}$ &  10.0 & 1.0 & 0.1 & 22.00 \\
\hline
\end{tabular}
\caption{Interface function definitions and their deviation scores.}
\label{tab:interface-functions}
\vspace{-0.3cm}
\end{table}

\bibliography{references}

@article{Bolhuis2002,
 author = {Bolhuis, Peter G and Chandler, David and Dellago, Christoph and Geissler, Phillip},
 title = {{Transition path sampling: {Throwing} ropes over mountain passes, in the dark}},
 journal = {Ann. Rev. Phys. Chem.},
 year = {2002},
 volume = {53},
 pages = {291--318},
}

@article{vanErp2003,
 author = {van Erp, Titus S and Moroni, Daniele and Bolhuis, Peter G},
 title = {{A novel path sampling method for the calculation of rate constants}},
 journal = {J. Chem. Phys.},
 year = {2003},
 volume = {118},
 number = {17},
 pages = {7762},
}

@article{Jung2023,
  title = {Machine-guided path sampling to discover mechanisms of molecular self-organization},
  volume = {3},
  ISSN = {2662-8457},
  url = {http://dx.doi.org/10.1038/s43588-023-00428-z},
  DOI = {10.1038/s43588-023-00428-z},
  number = {4},
  journal = {Nat. Comput. Sci.},
  publisher = {Springer Science and Business Media LLC},
  author = {Jung,  Hendrik and Covino,  Roberto and Arjun,  A. and Leitold,  Christian and Dellago,  Christoph and Bolhuis,  Peter G. and Hummer,  Gerhard},
  year = {2023},
  month = apr,
  pages = {334–345}
}

@article{vanerpReactionRateCalculation2007,
	title = {Reaction {Rate} {Calculation} by {Parallel} {Path} {Swapping}},
	volume = {98},
	copyright = {http://link.aps.org/licenses/aps-default-license},
	issn = {0031-9007, 1079-7114},
	url = {https://link.aps.org/doi/10.1103/PhysRevLett.98.268301},
	doi = {10.1103/PhysRevLett.98.268301},
	number = {26},
	urldate = {2024-11-13},
	journal = {Phys. Rev. Let.},
	author = {Van Erp, Titus S.},
	month = jun,
	year = {2007},
	pages = {268301},
}

@article{Dellago2002,
 author = {Dellago, Christoph and Bolhuis, Peter G. and Geissler, Phillip L.},
 journal = {Adv. Chem. Phys.},
 pages = {1--78},
 title = {Transition path sampling},
 volume = {123},
 year = {2002},
}

@article{bolhuisRareEventsMultiple2008,
  title = {Rare Events via Multiple Reaction Channels Sampled by Path Replica Exchange},
  author = {Bolhuis, Peter G.},
  year = {2008},
  month = sep,
  journal = {J. Chem. Phys.},
  volume = {129},
  number = {11},
  pages = {114108},
  issn = {0021-9606, 1089-7690},
  doi = {10.1063/1.2976011},
  urldate = {2024-11-13},
  langid = {english},
}

@article{leimkuhlerRationalConstructionStochastic2012,
  title = {Rational {{Construction}} of {{Stochastic Numerical Methods}} for {{Molecular Sampling}}},
  author = {Leimkuhler, B. and Matthews, C.},
  year = {2012},
  month = jun,
  journal = {Appl. Math. Res. X},
  pages = {34–56},
  issn = {1687-1200, 1687-1197},
  doi = {10.1093/amrx/abs010},
  urldate = {2024-11-13},
  langid = {english},
}

@article{Lechner2010,
 author = {Lechner, Wolfgang and Rogal, Jutta and Juraszek, Jarek and Ensing, Bernd and Bolhuis, Peter G.},
 doi = {10.1063/1.3491818},
 url = {https://doi.org/10.1063/1.3491818},
 year = {2010},
 month = nov,
 publisher = {{AIP} Publishing},
 volume = {133},
 number = {17},
 pages = {174110},
 title = {Nonlinear reaction coordinate analysis in the reweighted path ensemble},
 journal = {J. Chem. Phys.},
}

@article{ferrenbergOptimizedMonteCarlo1989,
  title = {Optimized {{Monte Carlo}} Data Analysis},
  author = {Ferrenberg, Alan M. and Swendsen, Robert H.},
  year = {1989},
  month = sep,
  journal = {Phys. Rev. Let.},
  volume = {63},
  number = {12},
  pages = {1195--1198},
  issn = {0031-9007},
  doi = {10.1103/PhysRevLett.63.1195},
  urldate = {2025-01-31},
  copyright = {http://link.aps.org/licenses/aps-default-license},
}

@article{swensonOpenPathSamplingPythonFramework2019,
  title = {{{OpenPathSampling}}: {{A Python Framework}} for {{Path Sampling Simulations}}. 1. {{Basics}}},
  shorttitle = {{{OpenPathSampling}}},
  author = {Swenson, David W. H. and Prinz, Jan-Hendrik and Noe, Frank and Chodera, John D. and Bolhuis, Peter G.},
  year = {2019},
  month = feb,
  journal = {J. Chem. Theor. Comp.},
  volume = {15},
  number = {2},
  pages = {813--836},
  issn = {1549-9618, 1549-9626},
  doi = {10.1021/acs.jctc.8b00626},
  urldate = {2024-11-13},
  copyright = {http://pubs.acs.org/page/policy/authorchoice\_ccbyncnd\_termsofuse.html},
  langid = {english},
}

@article{Torrie_1974,
 author = {Torrie, Glenn M. and Valleau, John P.},
 journal = {Chem. Phys. Lett.},
 number = {4},
 pages = {578},
 title = {{Monte} {Carlo} Free Energy Estimates Using Non-{Boltzmann} Sampling: {Application} to the Sub-Critical {Lennard}-{Jones} Fluid},
 volume = {28},
 year = {1974},
}

@article{Laio_2002,
 author = {Laio, Alessandro and Parrinello, Michele},
 title = {Escaping free-energy minima},
 journal = {Proc. Nat. Acad. Sci. USA},
 year = {2002},
 volume = {99},
 number = {20},
 pages = {12562},
 index = {},
}

@article{Dellago1997,
 author = {Dellago, Christoph and Bolhuis, Peter G and Csajka, F{\'e}lix S and Chandler, David},
 title = {{Transition path sampling and the calculation of rate constants}},
 journal = {J. Chem. Phys.},
 year = {1998},
 volume = {108},
 number = {5},
 pages = {1964--1977},
}

@article{Dellago2009,
 author = {Dellago, Christoph and Bolhuis, Peter G},
 journal = {Adv Polym Sci},
 title = {Transition Path Sampling and Other Advanced Simulation Techniques for Rare Events},
 pages = {167--233},
 volume = {221},
 year = {2009},
}

@article{Du2013,
 author = {Du, Wei-Na and Bolhuis, Peter G.},
 title = {{Adaptive single replica multiple state transition interface sampling}},
 journal = {{J. Chem. Phys.}},
 year = {{2013}},
 volume = {{139}},
 number = {{4}},
 doi = {{10.1063/1.4813777}},
 pages = {{044105}},
 issn = {{0021-9606}},
 eissn = {{1089-7690}},
 unique-id = {{ISI:000322949300008}},
}

@article{Rogal2010,
 author = {Rogal, Jutta and Lechner, Wolfgang and Juraszek, Jarek and Ensing, Bernd and Bolhuis, Peter G},
 title = {{The reweighted path ensemble}},
 journal = {J. Chem. Phys.},
 year = {2010},
 volume = {133},
 number = {17},
 pages = {174109},
}

@article{vanErp:2012fe,
 author = {van Erp, Titus S},
 title = {{Dynamical Rare Event Simulation Techniques for Equilibrium and Nonequilibrium Systems}},
journal = {Adv. Chem. Phys.},
volume = {151},
 year = {2012},
 pages = {27--60},
 publisher = {John Wiley {\&} Sons, Inc.},
 address = {Hoboken, NJ, USA},
 month = apr,
}

@article{vanErp2007,
 author = {van Erp, Titus},
 title = {{Reaction Rate Calculation by Parallel Path Swapping}},
 journal = {Phys. Rev. Lett.},
 year = {2007},
 volume = {98},
 number = {26},
 pages = {268301},
 month = jun,
}

@article{Bolhuis2008,
 author = {Bolhuis, Peter G},
 title = {{Rare events via multiple reaction channels sampled by path replica exchange}},
 journal = {J. Chem. Phys.},
 year = {2008},
 volume = {129},
 number = {11},
 pages = {114108},
}

@incollection{Bolhuis2009a,
 author = {Bolhuis, Peter G and Dellago, Christoph},
 year = {2009},
 title = {{Trajectory Based Molecular Rare Event Simulations}},
 publisher = {Wiley-VCH},
 address = {Hoboken},
 booktitle = {Reviews of Computational Chemistry},
}

@article{BolhuisDellago2015,
 author = {Bolhuis, P. G. and Dellago, C.},
 title = {{Practical and conceptual path sampling issues}},
 journal = {Eur. Phys. J. Spec. Top.},
 year = {{2015}},
 volume = {{224}},
 number = {{12}},
 pages = {2409-2427},
 doi = {{10.1140/epjst/e2015-02419-6}},
}

@article{vanErp2005,
 author = {van Erp, Titus S. and Bolhuis, Peter G.},
 journal = {J. Comput. Phys.},
 pages = {157--181},
 title = {Elaborating transition interface sampling methods},
 volume = {205},
 year = {2005},
}

@article{Du2014,
 author = {Du, Weina and Bolhuis, Peter G.},
 title = {{Sampling the equilibrium kinetic network of Trp-cage in explicit solvent}},
 journal = {J. Chem. Phys.},
 year = {{2014}},
 volume = {{140}},
 number = {{19}},
 doi = {{10.1063/1.4874299}},
 pages = {{195102}},
}

@article{Newton2015,
 author = {Newton, Arthur C and Groenewold, Jan and Kegel, Willem K and Bolhuis, Peter G},
 title = {Rotational diffusion affects the dynamical self-assembly pathways of patchy particles},
 journal = {Proc. Nat. Acad. Sci. USA},
 volume = {112},
 number = {50},
 pages = {15308--15313},
 year = {2015},
 publisher = {National Acad Sciences},
}

@article{vanErp2012,
 author = {van Erp, Titus S.},
 title = {{Dynamical rare event simulation techniques for equilibrium and nonequilibrium systems}},
 journal = {Adv. Chem. Phys.},
 year = {{2012}},
 volume = {{151}},
 pages = {27-60},
}

@article{Cabriolu2017,
 author = {Cabriolu, Raffaela and Refsnes, Kristin M. Skjelbred and Bolhuis, Peter G. and van Erp, Titus S.},
 doi = {10.1063/1.4989844},
 year = {2017},
 month = oct,
 publisher = {{AIP} Publishing},
 volume = {147},
 number = {15},
 pages = {152722},
 title = {Foundations and latest advances in replica exchange transition interface sampling},
 journal = {J. Chem. Phys.},
}

@article{bolhuis_transition_2021,
	title = {Transition {Path} {Sampling} as {Markov} {Chain} {Monte} {Carlo} of {Trajectories}: {Recent} {Algorithms}, {Software}, {Applications}, and {Future} {Outlook}},
	volume = {4},
	issn = {2513-0390, 2513-0390},
	shorttitle = {Transition {Path} {Sampling} as {Markov} {Chain} {Monte} {Carlo} of {Trajectories}},
	url = {https://onlinelibrary.wiley.com/doi/10.1002/adts.202000237},
	doi = {10.1002/adts.202000237},
	number = {4},
	urldate = {2023-02-02},
	journal = {Adc. Theor.  Simul.},
	author = {Bolhuis, Peter G. and Swenson, David W. H.},
	month = apr,
	year = {2021},
	pages = {2000237},
}

@article{Lazzeri2023,
  title = {Molecular Free Energies,  Rates,  and Mechanisms from Data-Efficient Path Sampling Simulations},
  volume = {19},
  ISSN = {1549-9626},
  url = {http://dx.doi.org/10.1021/acs.jctc.3c00821},
  DOI = {10.1021/acs.jctc.3c00821},
  number = {24},
  journal = {J.   Chem. Theor.  Comput.},
  publisher = {American Chemical Society (ACS)},
  author = {Lazzeri,  Gianmarco and Jung,  Hendrik and Bolhuis,  Peter G. and Covino,  Roberto},
  year = {2023},
  month = nov,
  pages = {9060–9076}
}

@article{Shirts2008,
  title = {Statistically optimal analysis of samples from multiple equilibrium states},
  volume = {129},
  ISSN = {1089-7690},
  url = {http://dx.doi.org/10.1063/1.2978177},
  DOI = {10.1063/1.2978177},
  number = {12},
  journal = {J. Chem. Phys.},
  publisher = {AIP Publishing},
  author = {Shirts,  Michael R. and Chodera,  John D.},
  year = {2008},
  month = sep 
}

@article{Kumar1992,
  title = {THE weighted histogram analysis method for free‐energy calculations on biomolecules. I. The method},
  volume = {13},
  ISSN = {1096-987X},
  url = {http://dx.doi.org/10.1002/jcc.540130812},
  DOI = {10.1002/jcc.540130812},
  number = {8},
  journal = {J.  Comput. Chem. },
  publisher = {Wiley},
  author = {Kumar,  Shankar and Rosenberg,  John M. and Bouzida,  Djamal and Swendsen,  Robert H. and Kollman,  Peter A.},
  year = {1992},
  month = oct,
  pages = {1011–1021}
}

@article{Arjun2020,
  title = {Rate Prediction for Homogeneous Nucleation of Methane Hydrate at Moderate Supersaturation Using Transition Interface Sampling},
  volume = {124},
  ISSN = {1520-5207},
  url = {http://dx.doi.org/10.1021/acs.jpcb.0c04582},
  DOI = {10.1021/acs.jpcb.0c04582},
  number = {37},
  journal = { J. Phys. Chem. B},
  publisher = {American Chemical Society (ACS)},
  author = {Arjun,  A. and Bolhuis,  P. G.},
  year = {2020},
  month = aug,
  pages = {8099–8109}
}

@article{Lechner2011,
  title = {Reaction coordinates for the crystal nucleation of colloidal suspensions extracted from the reweighted path ensemble},
  volume = {135},
  ISSN = {1089-7690},
  url = {http://dx.doi.org/10.1063/1.3651367},
  DOI = {10.1063/1.3651367},
  number = {15},
  journal = {J. Chem. Phys.},
  publisher = {AIP Publishing},
  author = {Lechner,  Wolfgang and Dellago,  Christoph and Bolhuis,  Peter G.},
  year = {2011},
  month = oct 
}

@article{vanErp2016,
  title = {Analyzing Complex Reaction Mechanisms Using Path Sampling},
  volume = {12},
  ISSN = {1549-9626},
  url = {http://dx.doi.org/10.1021/acs.jctc.6b00642},
  DOI = {10.1021/acs.jctc.6b00642},
  number = {11},
  journal = {J. Chem. Theor. Comput. },
  publisher = {American Chemical Society (ACS)},
  author = {van Erp,  Titus S. and Moqadam,  Mahmoud and Riccardi,  Enrico and Lervik,  Anders},
  year = {2016},
  month = oct,
  pages = {5398–5410}
}

@article{DazLeines2017,
  title = {Atomistic insight into the non-classical nucleation mechanism during solidification in Ni},
  volume = {146},
  ISSN = {1089-7690},
  url = {http://dx.doi.org/10.1063/1.4980082},
  DOI = {10.1063/1.4980082},
  number = {15},
  journal = {J. Chem. Phys.},
  publisher = {AIP Publishing},
  author = {Díaz Leines,  Grisell and Drautz,  Ralf and Rogal,  Jutta},
  year = {2017},
  month = apr 
}

@article{Liang2020,
  title = {Identification of a multi-dimensional reaction coordinate for crystal nucleation in Ni3Al},
  volume = {152},
  ISSN = {1089-7690},
  url = {http://dx.doi.org/10.1063/5.0010074},
  DOI = {10.1063/5.0010074},
  number = {22},
  journal = {J. Chem. Phys.},
  publisher = {AIP Publishing},
  author = {Liang,  Yanyan and Díaz Leines,  Grisell and Drautz,  Ralf and Rogal,  Jutta},
  year = {2020},
  month = jun 
}

@article{Newton2017,
  title = {The opposing effects of isotropic and anisotropic attraction on association kinetics of proteins and colloids},
  volume = {147},
  ISSN = {1089-7690},
  url = {http://dx.doi.org/10.1063/1.5006485},
  DOI = {10.1063/1.5006485},
  number = {15},
  journal = {J. Chem. Phys. },
  publisher = {AIP Publishing},
  author = {Newton,  Arthur C. and Kools,  Ramses and Swenson,  David W. H. and Bolhuis,  Peter G.},
  year = {2017},
  month = oct 
}

@article{Henin2022,
  title = {Enhanced Sampling Methods for Molecular Dynamics Simulations [Article v1.0]},
  volume = {4},
  ISSN = {2575-6524},
  url = {http://dx.doi.org/10.33011/livecoms.4.1.1583},
  DOI = {10.33011/livecoms.4.1.1583},
  number = {1},
  journal = {Living J.  Comput.  Mol. Sci. },
  publisher = {University of Colorado at Boulder},
  author = {Hénin,  Jérôme and Lelièvre,  Tony and Shirts,  Michael R and Valsson,  Omar and Delemotte,  Lucie},
  year = {2022}
}

@article{breebaart2025understandingreactionmechanismsstart,
      title={Understanding Reaction Mechanisms from Start to Finish}, 
      author={Breebaart, Rik S. and Lazzeri, Gianmarco and Covino, Roberto and Bolhuis, Peter G.},
      year={2025},
      eprint={2507.04052},
      archivePrefix={arXiv},
      primaryClass={physics.chem-ph},
      url={https://arxiv.org/abs/2507.04052}, 
}

@article{Bartels2000,
  title = {Analyzing biased Monte Carlo and molecular dynamics simulations},
  volume = {331},
  ISSN = {0009-2614},
  url = {http://dx.doi.org/10.1016/S0009-2614(00)01215-X},
  DOI = {10.1016/s0009-2614(00)01215-x},
  number = {5–6},
  journal = {Chem.  Phys. Lett.},
  publisher = {Elsevier BV},
  author = {Bartels,  Christian},
  year = {2000},
  month = dec,
  pages = {446–454}
}

@article{Safaei2025,
  title = {Exact Kinetics of Drug Permeation Using Transition Interface Sampling},
  volume = {129},
  ISSN = {1520-5207},
  url = {http://dx.doi.org/10.1021/acs.jpcb.5c05025},
  DOI = {10.1021/acs.jpcb.5c05025},
  number = {39},
  journal = { J.  Phys. Chem. B},
  publisher = {American Chemical Society (ACS)},
  author = {Safaei,  Sina and Baldauf,  Lukas and van Erp,  Titus S. and Ghysels,  An},
  year = {2025},
  month = sep,
  pages = {10019–10034}
}

@article{Zhang2024,
  title = {Highly parallelizable path sampling with minimal rejections using asynchronous replica exchange and infinite swaps},
  volume = {121},
  ISSN = {1091-6490},
  url = {http://dx.doi.org/10.1073/pnas.2318731121},
  DOI = {10.1073/pnas.2318731121},
  number = {7},
  journal = {Proc. Natl. Acad. Sci},
  publisher = {Proceedings of the National Academy of Sciences},
  author = {Zhang,  Daniel T. and Baldauf,  Lukas and Roet,  Sander and Lervik,  Anders and van Erp,  Titus S.},
  year = {2024},
  month = feb 
}

@book{FrenkelSmit,
  title     = "Understanding Molecular Simulation: From algorithms to  applications",
  author    = "Frenkel, Daan and Smit, Berend",
  publisher = "Elsevier",
  month     =  jul,
  year      =  2023,
  address   = "Amsterdam, Netherlands",
}

\end{document}